\documentclass[12pt]{JHEP3}

\usepackage{epsfig}
\epsfclipon
\usepackage{multicol}
\usepackage{epsfig,bm}
\usepackage{amssymb,amsmath}

\newcommand{\roughly}[1]{\mathrel{\raise.3ex\hbox{$#1$\kern-0.85em
\lower1ex\hbox{$\sim$}}}}

\newcommand{\lsim}{\roughly<}
\newcommand{\gsim}{\roughly>}

\def\be{\begin{equation}}
\def\ee{\end{equation}}
\def\bea{\begin{eqnarray}}
\def\eea{\end{eqnarray}}

\def\pref#1{(\ref{#1})}

\def\beq{\begin{equation}}
\def\eeq{\end{equation}}
\def\beqa{\begin{eqnarray}}
\def\eeqa{\end{eqnarray}}

\def\ssl{{\scriptscriptstyle L}}
\def\ssr{{\scriptscriptstyle R}}
\def\sst{{\scriptscriptstyle T}}

\def\cS{{\cal S}}
\def\cP{{\cal P}}

\def\hS{\hat{\cal S}}
\def\hP{\hat{\cal P}}

\newcommand{\bmat}{\left(\begin{array}}
\newcommand{\emat}{\end{array}\right)}

\def\nisubsubsection#1{\medskip\noindent{\bf #1}\\\smallskip\noindent}

\def\NPB#1#2#3{Nucl. Phys. B{#1} (19#2) #3}
\def\PLB#1#2#3{Phys. Lett. B{#1} (19#2) #3}

\def\PRD#1#2#3{Phys. Rev. D{#1} (19#2) #3}
\def\PRL#1#2#3{Phys. Rev. Lett. {#1} (19#2) #3}

\def\Journal#1#2#3#4{{#1} {\bf #2} (#4) #3}
\def\JETPUSSR{JETP (USSR)}
\def\ZETP{Zh. Eksp. Teor. Piz.}
\def\PTP{Prog. Theor. Phys.}
\def\PRD{Phys. Rev. D}
\def\EPJ{Euro. Phys. J. C}
\def\PRL{Phys. Rev. Lett.}
\def\NPB{Nucl. Phys. B}
\def\JHEP{JHEP}
\def\PLB{Phys. Lett. B}
\def\IJMP{Int. J. Mod. Phys. A}
\def\JPG{J. Phys. G}

\def\yzero{\smash{\hbox{$y\kern-4pt\raise1pt\hbox{${}^\circ$}$}}}

\def\-{\hphantom{-}}

\def\s2{\frac{1}{2}}

\def\IF{\relax{\rm I\kern-.18em F}}
\def\II{\relax{\rm I\kern-.18em I}}
\def\IP{\relax{\rm I\kern-.18em P}}
\def\IC{\relax{\rm I\kern-.48em C}}
\def\IR{\relax{\rm I\kern-.18em R}}
\def\IK{\relax{\rm I\kern-.20em K}}
\def\IM{\relax{\rm I\kern-.25em M}}

\def\Dsl{\,\raise.15ex\hbox{/}\mkern-13.5mu D} 
\def \one{\relax{\rm 1\kern-.26em I}}

\def\KK{{\scriptscriptstyle KK}}

\title{MSLED, Neutrino Oscillations\\ and the Cosmological Constant}

\author{J.~Matias$^1$ and C.P.~Burgess$^{2,3,4}$
\\

$^1$ IFAE, Universitat Aut\`onoma de Barcelona,
\\ \qquad 08193 Bellaterra, Barcelona, Spain.\\
${}^2$ Physics Department, McGill University, 3600 University
Street,\\ \qquad Montr{\'e}al, Qu{\'e}bec, Canada, H3A 2T8. \\
${}^3$ Department of Physics and Astronomy, McMaster University,\\
\qquad 1280 Main Street West, Hamilton, Ontario, Canada, L8S 4M1.\\
${}^4$ Perimeter Institute, 31 Caroline Street North,\\ \qquad
Waterloo, Ontario, Canada.\\
}

\date{}

\abstract{We explore the implications for neutrino masses and
mixings within the minimal version of the supersymmetric
large-extra-dimensions scenario (MSLED). This model was proposed
in {\tt hep-ph/0404135} to extract the phenomenological
implications of the promising recent attempt (in {\tt
hep-th/0304256}) to address the cosmological constant problem.
Remarkably, we find that the simplest couplings between brane and
bulk fermions within this approach can lead to a
phenomenologically-viable pattern of neutrino masses and mixings
that is {\it also} consistent with the supernova bounds which are
usually the bane of extra-dimensional neutrino models. Under
certain circumstances the MSLED scenario can lead to a lepton
mixing (PMNS) matrix close to the so-called bi-maximal or the
tri-bimaximal forms (which are known to provide a good description
of the neutrino oscillation data). We discuss the implications of
MSLED models for neutrino phenomenology.}


\begin{document}



\section{Introduction}

The supersymmetric Large Extra Dimensional (SLED) scenario was
proposed \cite{Towards,SLEDreviews} in an effort to provide the
long-sought understanding of why the vacuum energy should have
incredibly small value --- $\rho_{\rm vac} = \lambda^4$ with
$\lambda \sim 10^{-3}$ eV --- which is observed
\cite{DEdiscovery}. According to this proposal all observed
particles reside on a 3-brane situated within an extra-dimensional
world which has two large ($1/r \sim 10^{-2}$ eV)
dimensions\footnote{In our conventions the extra-dimensional
volume is $V_2 = r^2$, and so (for a square 2-torus, for example)
the Kaluza-Klein (KK) masses are of order $m_{\KK} \sim 2\pi/r$.}
(plus possibly others which are much smaller, $1/r \gsim 1$ TeV),
which can be possible \cite{ADD} provided the scale of gravity in
the large extra dimensions is of order the TeV scale.

The rationale for such a picture is that although
non-gravitational (brane) particles and interactions behave in the
usual way, the gravitational response to their zero-point vacuum
energy changes because gravity becomes 6-dimensional at energies
above $10^{-2}$ eV. Although the ultimate success of the proposal
is still under active study
\cite{SLEDrelated1,SLEDrelated2,SLEDrelated3}, its prospects for
success remain promising.

A particularly attractive feature of the SLED scenario is that it
is very predictive, with many testable implications for cosmology
and for tests of gravity on large and small scales. Better yet, it
has observable implications for high-energy collider experiments,
whose detailed implications have only recently begun to be
explored \cite{preSLEDPheno,SLEDPheno}. The predictiveness of the
SLED picture is most pronounced in its minimal version, MSLED
\cite{MSLED}, which assumes the particle content on our brane to
be given by precisely the Standard Model particle content.

Because of the similarity between the Kaluza-Klein (KK) scale in
these models, $2\pi/r \sim 10^{-1}$ eV, and the neutrino masses
which are observed in oscillation experiments, it is natural to
examine whether these models might also have observational
implications for neutrino physics. We examine this issue here, and
find a rich set of phenomena can arise within the neutrino sector.
In particular, we argue that the observed pattern of neutrino
masses and mixings can emerge through the mixing of the
brane-bound neutrinos with the many massless 6D fermions which
SLED models predict must appear in the bulk, without running afoul
of the bounds on sterile neutrinos.

\subsection{Neutrino Masses in MSLED}

The possibility that large extra dimensions might have
implications for neutrino masses has been explored with many
variations
\cite{LEDneutrinos,LEDneutrinos2,LEDneutrinos3,Rabi,Keith20001}.
The vast majority of these variations share the common feature
that the neutrino mass eigenstates arise when the usual
brane-bound neutrino flavours mix with various higher-dimensional
neutral fermions which are assumed to reside within the bulk. It
is generically assumed that these bulk fermions are massless in
the higher-dimensional sense, and so from the 4-dimensional point
of view only acquire masses (before mixing with the brane
neutrinos) through the Kaluza-Klein mechanism. Various models are
obtained by adjusting the strength of the brane-bulk couplings,
the number of extra dimensions and the size of each dimension.

We follow a similar construction here, but find that the results
from supersymmetric large extra dimensions are much more
predictive, for the following reasons.
\begin{itemize}
\item The massless field content in the bulk is dictated by
6-dimensional supersymmetry, and automatically contains numerous
candidate neutral fermion fields which can mix with the brane
neutrinos. Better yet, there is an understanding why the 6D mass
of these bulk fermions must vanish. They vanish because they are
related by supersymmetry to massless bosons, such as the graviton.
\item There is little freedom to alter the size and shape of the
large extra dimensions, because the value of the observed Dark
energy density dictates that there are two large dimensions, and
the spacing of KK masses in {\it both} of these dimensions is not
larger $m_{\KK} \sim 2\pi/r \sim 0.1$ eV.
\item Like for other models with large extra dimensions, there is
a natural extension of lepton number to extra-dimensional fermions
--- chirality in the two large dimensions --- which is also
naturally broken at the scale $m_{\KK}$. This allows a technically
natural understanding of why quantum corrections do not
destabilize the observed neutrino mass pattern.
\item The back-reaction of the branes onto the bulk geometry is
important in SLED models because it is part of the mechanism for
suppressing the vacuum energy. In particular, this is how the bulk
learns about supersymmetry breaking on the branes because the
boundary conditions which the branes set up for the bulk fermions
typically remove the massless fermionic Kaluza-Klein modes (like
the gravitini) from the spectrum. This absence of bulk-fermion
zero modes also removes the massless modes from the bulk fermions
which mix with the brane neutrinos. This is important
phenomenologically, since it is the degeneracy (in the absence of
brane-bulk couplings) between these massless bulk states with the
massless brane states which typically leads to large mixings
between these states once brane-bulk couplings are turned on. In
this way SLED models sidestep one of the phenomenological
difficulties of large-extra-dimensional neutrino
models.\footnote{Back-reaction issues are included in the more
detailed microscopic construction of ref.~\cite{LEDneutrinos3}.}
\end{itemize}
Remarkably, we find that this restrictiveness can nonetheless very
naturally lead to a phenomenologically successful picture of
neutrino mixing.

\subsection{Required Neutrino Properties}

Before describing the neutrino mass and mixing pattern to which
SLED models lead, we first pause to summarize the main results of
current neutrino phenomenology which must be explained.

\subsubsection*{Active Neutrino Oscillations}

The observed pattern of neutrino oscillations are consistent with
there being just the 3 active neutrino species, whose
participation in the charged-current weak interactions has the
form
\begin{equation} \label{CCInteraction}
    {\cal L}_{cc} = \frac{ig \, U_{ai} }{\sqrt2} \;
    W_\mu \, (\overline{\ell}_a \gamma^\mu \,
    \gamma_\ssl \, \nu_i) + \hbox{h.c.} \,,
\end{equation}
where $\gamma_\ssl$ is the projector onto left-handed spinors and
$i = 1,2,3$ labels the 3 neutrino types, as $a = 1,2,3$ does for
the leptons: $\{ \ell_1, \ell_2 , \ell_3 \} = \{ e, \mu , \tau
\}$. Here the {\sl PMNS} matrix \cite{PMNS}, $U_{ai}$, describes
the amplitude with which the neutrino type `$i$' reacts with the
charged lepton `$a$'. For three types of neutrinos it may be
parameterized in terms of mixing angles and phases: $U = V \, K$,
with
\begin{eqnarray}
    V &=&\left( {\begin{array}{*{20}c}
   1 & 0 & 0  \\
   0 & {c_{23} } & {s_{23} }  \\
   0 & { - s_{23} } & {c_{23} } \\
    \end{array}} \right)
    \left( {\begin{array}{*{20}c}
   {c_{13} } & 0 & {s_{13}e^{-i\delta} }  \\
   0 & 1 & 0  \\
   { - s_{13}e^{i\delta} } & 0 & {c_{13} }  \\
    \end{array}} \right)
    \left( {\begin{array}{*{20}c}
   {c_{12} } & {s_{12} } & 0  \\
   { - s_{12} } & {c_{12} } & 0  \\
   0 & 0 & 1  \\
    \end{array}} \right)
    \nonumber \\
    &=&\left( \begin{array}{ccc}
  c_{12}c_{13} &  s_{12}c_{13}&  s_{13}e^{-i\delta}\\
  -c_{23}s_{12}-s_{23}c_{12}s_{13}e^{i\delta}
                                 &  c_{23}c_{12}-s_{23}s_{12}s_{13}e^{i\delta}
                                 &  s_{23}c_{13}\\
  s_{23}s_{12}-c_{23}c_{12}s_{13}e^{i\delta}
                                 &  -s_{23}c_{12}-c_{23}s_{12}s_{13}e^{i\delta}
                                 & c_{23}c_{13}\\
    \end{array} \right)
    \\\label{Eq:U_nu}
    \hbox{and}\qquad
    K &=& \begin{pmatrix} e^{i\rho} && \\ & e^{i\sigma} & \\
    && 1 \end{pmatrix} \,, \label{Eq:K}
\end{eqnarray}
where $c_{ij}=\cos\theta_{ij}$ and $s_{ij}=\sin\theta_{ij}$.

As is well known, assuming only three neutrinos to be relevant
implies the splitting between two of these neutrinos must be much
smaller than their common splitting from the third, and we follow
common practice by choosing our labelling for the neutrino
eigenstates such that the small splitting is between $\nu_1$ and
$\nu_2$. With this choice, the successful description of the
oscillation lengths seen in solar- and atmospheric-neutrino
experiments implies the three neutrino masses, $m_{1,2,3}$, must
satisfy \cite{RecentData,RecentAnalyses}
\begin{eqnarray}
    \Delta m^2_{atm} &=& \vert m^2_3-m^2_2\vert = (1.1-3.4)\times
    10^{-3}~{\rm eV}^2, \nonumber\\ \Delta m^2_\odot &=& \vert
    m^2_2-m^2_1\vert = (5.4-9.4)\times 10^{-5}~{\rm eV}^2 \,.
\end{eqnarray}
This range for solar neutrinos corresponds to the
Large-Mixing-Angle (LMA) solution, which is the only one which
accounts for the most recent results from the SNO and KamLAND
oscillation measurements. In themselves neutrino oscillations do
not fix the values of each of the masses separately, leaving open
two possibilities. If the degenerate pair, $\nu_1$ and $\nu_2$,
are less massive than $\nu_3$, then the mass pattern is known as a
`normal' hierarchy, while if $m_1, m_2
> m_3$ the hierarchy is `inverted'.

With this choice, the absence of the observation of short-distance
oscillations of reactor-generated neutrinos implies the 3$\sigma$
limit
\begin{equation}
     \vert U_{e3}\vert^2 = s^2_{13} < 0.067 \,.
    \label{Eq:Data13}
\end{equation}
Because this implies $c_{13} > 0.966$, we have $1 > |U_{e1}|^2 +
|U_{e2}|^2 = c_{13}^2 > 0.933$ and so it is convenient to write
$|U_{e1}| = c_{12} c_{13} \simeq \cos \theta_\odot$ and $|U_{e2}|
= s_{12} c_{13} \simeq \sin\theta_\odot$ when describing other
oscillation measurements. Observations of solar-neutrino
oscillations constrain $s_{12}$, and imply
\begin{eqnarray}
    0.67\leq \sin^22\theta_\odot \leq 0.93 \,,
    \label{Eq:Data12}
\end{eqnarray}
where, as above, $\theta_\odot \simeq \theta_{12}$. Oscillations
as seen in atmospheric neutrinos are close to maximal, and satisfy
\begin{eqnarray}
    0.85 \le \sin^22\theta_{\rm atm}
    \le 1 \,,
    \label{Eq:Data23}
\end{eqnarray}
where $\theta_{\rm atm} = \theta_{23}$.

\medskip
\nisubsubsection{Suggestive Textures}
For later convenience it is worth remarking that a good first
description for the observed values of the elements of the PMNS
matrix is obtained by taking $\theta_{13} \approx 0$ and so
$s_{13} \approx 0$, $c_{13} \approx 1$. We may also approximate
atmospheric-neutrino mixing as being maximal, in which case
$\theta_{23} \approx {\pi}/{4}$ and $s_{23} \approx c_{23} \approx
{1}/{\sqrt2} = 0.707..$, so $\sin^22\theta_{\rm atm} =
\sin^22\theta_{23} = 1$. With these choices the mixing matrix
takes the form
\begin{equation} \label{GoodUMatrix}
    U \approx
    \left( \begin{array}{ccccc} c_{12} &&  s_{12} &&  0\\
    -s_{12}/\sqrt2 &&  {c_{12}}/\sqrt2 &&  1/\sqrt2 \\
    s_{12}/\sqrt2 &&  - c_{12}/\sqrt2 && 1/\sqrt2 \\
    \end{array} \right) \,.
\end{equation}

These choices are easy to obtain within specific models like the
extra-dimensional ones to be discussed shortly, because they
follow as consequences of the assumption that the masses are
invariant under a discrete $Z_2$ symmetry
\cite{z2sym,z2sym2,z2sym3} which interchanges the second and third
generations. To see this, notice that the most general $3\times 3$
mass matrix which is invariant under this interchange may be
written
\begin{equation}
    M_{Z_2} =
    \left( \begin{array}{cccccc} m &&&  m' &&  m' \\
    m' &&&  m + \hat{m} &&  \hat{m} \\
    m' &&&  \hat{m} && m + \hat{m} \\
    \end{array} \right) \,,
\end{equation}
and this matrix always admits the eigenvector $\nu \propto
(0,1,-1)^\sst$ with eigenvalue $m$. As we shall see, imposing this
symmetry on extra-dimensional models gives a matrix of this form,
but with $m = 0$, leading to an inverted-hierarchy mass pattern
for which $\nu_3$ is massless.

In later sections we shall be led towards two particular choices
of mixing matrices whose entries are very similar (up to signs) to
the form of eq.~\pref{GoodUMatrix}. For one case $s_{12} \approx
-1/\sqrt3 = -0.577...$ and $c_{12} \approx \sqrt{2/3} = 0.816...$
(and so $\sin^2 2\theta_{12} = 4 s_{12}^2 c_{12}^2 = 8/9 =
0.888...$), which is called the tri-bimaximal-mixing form
\cite{MaximalMixing}, and lies within the observed range allowed
by oscillation measurements. In this case the PMNS matrix becomes
\begin{equation} \label{TriBiMaximal}
    U \approx
    \left( {\begin{array}{
    ccccc}
    \sqrt{2/3} && -1/{\sqrt3} &&
    0  \\
    1/\sqrt6 && 1/{\sqrt3} &&
    1/{\sqrt2}  \\
    -1/\sqrt6 && - 1/{\sqrt3} &&
    1/{\sqrt2} \\
    \end{array}} \right) \,.
\end{equation}

A second example to which later sections lead is the bi-maximal
mixing form, for which $-s_{12} = c_{12} = 1/\sqrt2$, and so for
which the PMNS matrix becomes
\begin{equation} \label{BiMaximal}
    U \approx
    \left( {\begin{array}{
    ccccc}
    {1/\sqrt2} && -1/{\sqrt2} &&
    0  \\
    1/2 && 1/2 &&
    1/{\sqrt2}  \\
    -1/2 && - 1/2 &&
    1/{\sqrt2} \\
    \end{array}} \right) \,.
\end{equation}
This form does {\it not} successfully describe the data, because
the maximal value $\sin^22\theta_\odot = 4 s_{12}^2 c_{12}^2 = 1$
lies outside of the experimentally-allowed range. However it
naturally arises if $\nu_1$ and $\nu_2$ are `pseudo-Dirac', or
carry an approximately-conserved lepton number \cite{PseudoDirac}.
In this case it is the perturbations which break the relevant
lepton number which change $s_{12}$ and bring the resulting PMNS
matrix into agreement with observations.

\subsubsection*{Bounds on Individual Sterile Neutrinos}

Any 4D fermion which transforms as a singlet under the Standard
Model gauge group can become a sterile neutrino if it mixes with
any of the three active neutrinos. If such mixings exist then the
PMNS matrix, $U_{ai}$, for the charged-current interactions is no
longer square, since it has 3 rows ($a=1,2,3$, for each charged
lepton) but $3+N$ columns ($i=1,...,3+N$ if the 3 active neutrinos
are supplemented by $N$ sterile counterparts). By virtue of these
new mixing matrix elements sterile neutrinos can have observable
signatures, none of which have been observed to date. This section
summarizes the observational bounds which follow from these
potential signatures.

Many of the phenomenological constraints on sterile neutrinos are
performed under the minimal assumption that only a single species
of sterile neutrino exists ({\it i.e.} $N=1$)
\cite{SingleSterile,SNOSterile,SNFeedback1}. The bounds which
result typically rely on the absence of oscillations between this
neutrino and the usual active ones, and so the strength of the
allowed mixing can depend sensitively on what is assumed about the
mass difference, $\Delta m^2_{14}$, between the sterile and the
relevant active neutrino state and about the active-sterile mixing
parameter, $\varepsilon \sim |\sin\theta_s| \ll 1$.\footnote{The
presence of sterile neutrinos can also modify non-neutrino
precision electroweak measurements, such as by modifying the muon
lifetime from which the Fermi constant is inferred \cite{BigFit},
although these are not yet competitive with more direct
observables.} We summarize some of these bounds here, and return
in later sections to describe the somewhat stronger bounds which
arise for the higher-dimensional models of present interest.

\begin{itemize}
\item {\it Solar Neutrinos:} The agreement between solar-model
predictions and the observed charged-current and neutral-current
solar-neutrino fluxes at SNO constrain the amount of flux which
can be lost to sterile states. For generic neutrino mass
differences larger than $10^{-12}$ eV${}^2$ these bounds constrain
the sterile-active neutrino mixing angle to be smaller than
$\varepsilon \lsim 0.1$, where the precise bound depends on the
mass difference. This bound improves to $\varepsilon \lsim 0.001$
for mass splittings in a narrow range around $\Delta m^2_{14} =
10^{-4}$ eV${}^2$, and to $\varepsilon \lsim 0.01$ for $10^{-8}
\hbox{eV}^2 \lsim \Delta m_{14}^2 \lsim 10^{-6}$ eV${}^2$, due to
the presence in these cases of resonant active-sterile MSW
oscillations within the Sun.
\item {\it Reactor Neutrinos:} The absence of neutrino flux
disappearance from reactor-generated neutrinos constrains
$\varepsilon \lsim 0.1$ for $\Delta m_{14}^2 \gsim 10^{-3}$
eV${}^2$.
\item {\it Atmospheric Neutrinos:} Active-sterile neutrino
oscillations can be excluded for $\varepsilon \lsim 0.2$ for the
range $10^{-4} \lsim \Delta m_{14}^2 \lsim 10^{-2}$ eV${}^2$.
\item {\it Supernova Neutrinos:} Excessively large active-sterile
anti-neutrino conversions within supernovae can drain flux from
the supernova neutrino signal, and too much of this can be
excluded from the observed signal from SN1987a. For resonant
$\nu_e-\nu_s$ oscillations, this can constrain mixings down to
$\varepsilon \lsim 0.01$ for mass splittings, $\Delta m_{14}^2$,
which are larger than around 10 eV${}^2$. However because neutrino
densities themselves play a significant role in these resonant
oscillations, there is a feedback mechanism through which
efficient resonant active-sterile oscillations can change the
neutrino densities and thereby rapidly turn off the resonance
\cite{SNFeedback1,SNFeedback2}.
\item{\it Nucleosynthesis:} The dominant cosmological constraint
for sterile neutrinos whose masses are smaller than 1 eV comes
from the requirement that there not be too many light degrees of
freedom contributing to the universal expansion during Big-Bang
Nucleosynthesis (BBN). Even if the primordial abundance of sterile
neutrinos is small, this constrains the strength of active-sterile
oscillations because these can cause too efficient production of
sterile neutrinos from the equilibrated active neutrinos. This
latter condition can constrain neutrino abundances down to mass
differences of order $10^{-8}$ eV${}^2$, with the strongest
constraints (coming for the largest mass differences) being of
order $\varepsilon \lsim 0.1$.
\item {\it Cosmic Microwave Background:} Measurements of CMB
temperature fluctuations constrain the energy density which can be
present in neutrinos at the epoch of recombination, leading to the
constraint $\Omega_\nu h^2 < 10^{-2}$. Since this constrains the
energy density present in neutrinos, the mixing angles to which
this allows constraints are larger for heavier neutrinos, being
sensitive to mixing angles as small as $\varepsilon \sim 0.001$
for sterile neutrino masses of order 10 eV.
\end{itemize}

\medskip\nisubsubsection{Bounds on Extra-Dimensional Sterile
Neutrinos}
Since large-extra-dimensional models contain entire KK towers of
sterile neutrinos, much more restrictive bounds are possible. For
the purposes of discussing these bounds we anticipate the next
section's results, and regard the sterile states to have masses
$m_\ell \approx c_\ell/r$, and mixings $\varepsilon_\ell$ with
active neutrino flavours, where $\ell$ labels the modes and
$c_\ell$ and $\varepsilon_\ell$ are dimensionless parameters which
depend on the details of the compactification. As we shall see
from explicit diagonalizations, typically $\varepsilon_\ell \sim
g/c_\ell$ where $g$ is a constant which is independent of $\ell$.

The bounds in this case come in two forms. First, the presence of
enormous numbers of sterile KK modes having masses ranging upwards
from $m_{KK} \sim 2\pi/r$, opens the possibility of there being
active-sterile neutrino oscillations into specific sterile states
for a great variety of oscillation lengths. This allows the use of
the best of the bounds listed above for individual sterile
neutrinos, provided only that the relevant mass differences are
included amongst the allowed KK neutrino masses.

The second kind of bound relies on the enormous available phase
space which the extra dimensions make available, since this allows
bounds to be obtained from the absence of processes which radiate
energy into sterile neutrinos {\it incoherently}, rather than
through coherent active-sterile neutrino oscillations. These
incoherent neutrino bounds are in addition to the similar bounds
which constrain the amount of incoherent emission which may be
tolerated into other species of bulk particles, such as gravitons.
A closely related class of constraints come from the cosmological
limits on the total energy which can be tied up at present in
sterile neutrinos. These bounds are particularly potent for
extra-dimensional models, for which enormous numbers of massive
and stable sterile neutrino states can be present.

Bounds of these second type are normally used to completely
exclude the possibility of explaining neutrino masses in terms of
more than a single large extra dimension, since each additional
large dimension enormously increases the amount of phase space
which is available in bulk neutrino states
\cite{LEDNuBounds,Raffelt,SNSterile,Barbieri,LEDCosmo,SNOSterileLED}.
The main novelty of our results in this paper is our ability to
provide a phenomenologically viable description of neutrino masses
using {\it two} large extra dimensions without running afoul of
these energy-loss bounds.

Amongst the most restrictive new bounds obtained in this way are

\begin{itemize}
\item {\it Supernova Bounds:} The large number of KK states makes
energy-loss bounds from supernovae very strong. Each KK mode is
radiated incoherently ({\it i.e.} by vacuum oscillations) with
rate $\Gamma_i \sim (m/E)^2 \Gamma_\nu$, where $m \sim g/r$ is a
measure of the mass term which mixes the active and sterile
neutrinos (see the next section for precise definitions), $E$ is
the neutrino energy and $\Gamma_\nu$ is a typical active-neutrino
emission rate. Since the number of sterile neutrino states having
mass smaller than $E$ is of order $(Er/2\pi)^n$ for $n$ extra
dimensions, an estimate for the total incoherent emission rate
into sterile states for $n=2$ is
\beq
    \Gamma_{\rm inv} = \sum_i \Gamma_i \sim \left( \frac{g}{rE}
    \right)^2 \Gamma_\nu \left( \frac{Er}{2\pi} \right)^2
    \sim \left( \frac{g}{2\pi} \right)^2 \Gamma_\nu \,.
\eeq
Refs.~\cite{Raffelt,Barbieri} require this to be at most $10^{-8}$
of the active-neutrino rate, $\Gamma_\nu$, which leads to the
conservative bound on the dimensionless bulk-brane mixing of order
$g / 2\pi \lsim 10^{-4}$.

We call this a conservative bound because it relies fairly
strongly on the present-day understanding of supernova explosions,
which are arguably relatively poorly understood given the
inability of current computer models to successfully explode a
star.\footnote{We thank Marco Cirelli and George Fuller for
helpful conversations on this point.} Another approach is only to
require that the rate of sterile-neutrino radiation not be more
than of the same order as the active-neutrino signal, in which
case only a comparatively weak constraint on $g$ is obtained.

For supernovae, another potentially very dangerous mechanism for
energy loss into sterile neutrinos is possible because the large
number of neutrino states allows a succession of potential
resonant oscillations which can depress the active-neutrino
survival rate as it passes through the supernova environment. For
small mixing each resonance is narrow and the survival probability
can be understood as the product of that for each KK mode. Even
though each KK mode mixes with an amplitude which is suppressed by
the mode's mass, the resonance for each has essentially the same
adiabaticity parameter, leading to a prohibitively large
conversion rate even for extremely small brane-bulk mixing
parameter $g$. Happily, as pointed out in
refs.~\cite{SNFeedback1,SNFeedback2}, this constraint is too naive
since the feedback of such oscillations onto the supernova
neutrino densities is likely to quickly reduce the rate of energy
loss in this channel to the vacuum-oscillation rate described
above.
\item {\it Late-Epoch Cosmology:} Large-extra-dimensional models
face well-known dangers with cosmology since the various KK states
for the many bulk fields can cause problems with Big Bang
Nucleosynthesis or with over-closure of the universe if they are
too abundant \cite{LEDCosmo}. These bounds can be avoided if the
KK modes do not appreciably decay into photons and if they and
their decay products are not too abundant at the BBN epoch.
Although it remains a challenge to obtain a pre-BBN cosmology
which gives sufficiently few relic KK modes in models with large
extra dimensions, we put this issue aside here since its proper
understanding must also await a hitherto missing study of the
time-dependence of 6D cosmological background solutions. We here
therefore follow the usual practice and assume there to be an
acceptable number of KK modes at BBN, and focus specifically on
the neutrino-related constraints which arise due to the
possibility of populating KK sterile neutrinos through
active-sterile oscillations from the known equilibrium abundance
of active neutrinos.

Constraints on the present-day energy density in neutrinos
preclude there being a significant number of stable massive
sterile neutrino states having masses greater than of order 1 eV
and having an appreciable mixing with active neutrinos. Mixings
are dangerous in this context because they can keep sterile
neutrinos in equilibrium down to low temperatures, leading to too
large late-time abundances. An efficient way to evade these bounds
is to require all of the sterile neutrinos to thermally decouple
at a temperature, $T_D$, above the QCD phase transition so their
abundance can be diluted by the entropy release which occurs
during this transition. This requires $T_D \gsim 1$ GeV, where for
each sterile neutrino species $T_D$ is related to the
active-sterile mixing parameter $\varepsilon$ by the condition
\beq
    \varepsilon^2 G_F^2 T^5_D \sim \frac{T^2_D}{M_p} \,,
\eeq
where $G_F \sim 10^{-5}$ GeV${}^{-2}$ is the Fermi coupling and
$M_p \sim 10^{18}$ GeV is the Planck mass. The condition $T_D \sim
1/(\varepsilon^2 G_F^2 M_p)^{1/3} \gsim 1$ GeV leads to the
constraint $\varepsilon \lsim 10^{-4}$. In later sections we find
$\varepsilon_\ell \sim g/c_\ell$, where $g$ measures the
dimensionless brane-bulk coupling and $c_\ell/r \gsim 2\pi/r$ is
the mass of the KK mode in question. We see that the constraint is
strongest for the lightest modes, and for modes with $c_\ell \sim
2\pi$ this constraint implies $g \lsim 2 \pi \times 10^{-4}$.
\end{itemize}

We see that both of the above constraints may be satisfied
provided that the brane-bulk couplings satisfy $g \lsim 2\pi
\times 10^{-4}$, and for this reason we have the regime of small
$g$ in mind when we describe the neutrino mixings in more detail
in the next section.

\section{Neutrinos in MSLED}

In MSLED neutrino masses can arise through mixing between the
usual three brane-bound neutrinos, $\nu_a$, and fermions in the
bulk. In order to describe this mixing we first digress to
summarize our conventions for 6D spinors.

\subsection{6D Spinors}

In 6 dimensions the Dirac matrices are $8 \times 8$ matrices,
which we take to have the following form: $\Gamma_M =
\{\Gamma_\mu, \Gamma_m \}$, where $\mu = 0,1,2,3$ and $m = 4,5$
and
\beq
    \Gamma_\mu = \gamma_\mu \otimes I_2 \qquad \hbox{and}
    \qquad \Gamma_m = \gamma_5 \otimes \tau_m \,.
\eeq
Here $\gamma_\mu$ denotes the usual $4 \times 4$ Dirac matrices,
while
\beq
    \tau_4 = \begin{pmatrix} 0 & -i \\ i & 0 \end{pmatrix}
    \qquad \hbox{and} \qquad
    \tau_5 = \begin{pmatrix} 0 & 1 \\ 1 & 0 \end{pmatrix} \,.
\eeq
In addition we define the usual 4- and 2-dimensional chirality
projectors $\gamma_\ssl = \frac12 (1 + \gamma_5)$, $\gamma_\ssr =
\frac12 (1 - \gamma_5)$ and $\tau_\pm = \frac12 (1 \pm \tau_3)$,
with
\beq
    \tau_3 = \begin{pmatrix} 1 & 0 \\ 0 & -1 \end{pmatrix} \,.
\eeq
In terms of these matrices, 6D chiral spinors are distinguished by
the eigenvalues of the matrix $\Gamma_7 = \gamma_5 \otimes
\tau_3$.

With these choices 6D spinors carry 4D and 2D indices,
$(u,\alpha)$, where $u = 1,...,4$ and $\alpha = 1,2$. It is useful
to take the basis for these spinors to have definite chirality,
since a 6D spinor with positive chirality, $\Gamma_7 \psi = +
\psi$, has 4 complex components and decomposes into two 4D states
for which the eigenvalues of the pair $(\gamma_5,\tau_3)$ are
$(\lambda,s) = (+1,+1)$ and $(\lambda,s) = (-1,-1)$.

Once compactified, we see that each 6D fermion reduces to a
Kaluza-Klein tower of 4D fermions. For example a collection of 6D
$\Gamma_7 = +1$ Weyl fermions has a 4D expansion of the form
\beq
    N^I(x,y) = \frac{1}{r} \sum_{s = \pm}
    \sum_\ell  u_{\ell s}(y) \, n^{I}_{\ell s}(x) \,,
\eeq
where $\ell$ labels the KK modes, with $u_{\ell s}(y)$ being
2-component spinors which are chosen to satisfy $\tau_3 \, u_{\ell
s} = s \, u_{\ell s}$. Consequently (because $\Gamma_7 N^I = +
N^I$) the 4D spinors $n^I_{\ell s}$ have 4D chirality $\gamma_5 \,
n^I_{\ell s} = s \, n^I_{\ell s}$ (notice that in the above all
spinor indices are suppressed). An overall factor of the size of
the extra dimensions, $1/r$, has been extracted from the
extra-dimensional fermion wave-functions, $u_{\ell s}(y)$, so that
these can satisfy $r$-independent orthonormality relations.

\subsection{Bulk-Brane Couplings}

The lowest-dimension bulk-brane interaction between left-handed
lepton fields on the branes and a collection of $\Gamma_7 = +1$ 6D
fermions, $N^I$, has the form
\beq \label{bulkbranenumixing}
    S_{{\rm int},N} =  \int d^4x \;
    \Bigl[ {\lambda_{aI\alpha}}
    (L_{a}^u N^I_{u\alpha})
    H + \hbox{c.c.} \Bigr] \,.
\eeq
Here the $\lambda_{aI\alpha}$ are coupling constants having
dimensions of inverse mass, which we expect to be roughly of order
$M_g^{-1}$ in size, where $M_g \sim 10$ TeV \cite{MSLED} is the
gravity scale in the bulk. In this expression the $SU_L(2)$ gauge
index is suppressed, while the index $a = 1,2,3$ labels fermion
generations. Similarly $u = 1,...,4$ and $\alpha = 1,2$ are the 4D
and 2D spinor indices, which we here temporarily re-instate.

It is convenient to write the couplings in terms of definite 2D
chirality, with
\beq
    \lambda_{aI\alpha} = \sum_{s = \pm} \lambda^{(s)}_{aI} \,
    u^{(s)\dagger}_\alpha \,,
\eeq
with $\tau_3 \, u^{(s)} = s \, u^{(s)}$. If both
$\lambda^{(+)}_{aI}$ and $\lambda^{(-)}_{aI}$ are nonzero for some
choice for $I$ and $a$ then these brane-bulk couplings violate the
2D local Lorentz transformations generated by
\beq
    J = \frac{i}{4} \, [\Gamma^4, \Gamma^5]
    = \frac{i}{4} \, \Bigl( I \otimes [\tau_4,\tau_5] \Bigr)=
    \frac{1}{2} \, \Bigl( I
    \otimes \tau_3 \Bigr) \,,
\eeq
which act on the 2D spinor index $\alpha$. On the other hand, if
all of the $\lambda^{(+)}_{aI}$ (or $\lambda^{(-)}_{aI}$) should
vanish, then the coupling preserves this $O(2)$ symmetry provided
that the brane fermions are also taken to be charged under it. Any
such $O(2)$ symmetry would be naturally interpreted as a
lepton-number symmetry for the neutrino mass matrix, since this is
the only possible way such a symmetry can be extended to an
invariance of the rest of the Standard Model brane couplings given
the assumption of minimal particle content.

Models with bulk-brane neutrino mixing have been studied in detail
in ref.~\cite{LEDneutrinos,LEDneutrinos2,LEDneutrinos3,Rabi}, but
this earlier round of model building differs from the present
picture in several important ways.

\medskip\noindent$\bullet$ {\bf Brane Back-Reaction:}
First, these earlier analyses ignore the brane back-reaction onto
the bulk. However, we know quite generally in MSLED that the
back-reaction of the brane (or branes) onto the bulk geometry
induces boundary conditions for the bulk fermions at the brane
position, and these remove the otherwise-massless Kaluza-Klein
level \cite{Towards}. (Indeed this is how the bulk modes in MSLED
`see' that supersymmetry is broken by the branes.)

\medskip\noindent$\bullet$ {\bf Approximate Lepton Number
Conservation:} We have seen that the brane-bulk mixing relates
lepton number on the brane to local $O(2)$ Lorentz transformations
in the bulk. Because of this connection it is guaranteed to be a
symmetry of the interactions of the 6D supergravity in the bulk.
In general this symmetry is spontaneously broken by the background
geometry, since extra-dimensional Lorentz rotations are always
broken by the zweibein of the internal two dimensions, ${e_m}^a$.
For some geometries (such as a 2-sphere) it can happen that the
transformation of ${e_m}^a$ under such Lorentz transformations can
be compensated by performing a diffeomorphism in the extra
dimensions, provided that the extra-dimensional geometry admits a
rotational isometry. In such cases the background breaks the
product of local Lorentz transformation and diffeomorphism
symmetries down to the diagonal rigid rotation which preserves
${e_m}^a$, in an extra-dimensional analogue of the spin-orbit
coupling. For geometries without rotational isometries (such as
2-tori) no such compensation is possible and the symmetry is
broken completely. This leads to a naturally very small bulk
lepton-number breaking scale, of order the KK mass scale, $m_{\KK}
\sim 0.1$ eV. Because of this low symmetry-breaking scale, it is
technically natural to have the scale of lepton-number violation
in the bulk-brane couplings also be as small as $m_{\KK}$.

\medskip\noindent$\bullet$ {\bf Bulk SUSY:}
In previous studies the properties of the bulk fermions were
usually adopted for the convenience of neutrino phenomenology, for
lack of a theory of the origin of the bulk fermions, $N^I$. In
MSLED, however, supersymmetry strongly constrains the properties
of the bulk. There are, for instance, fermions which are universal
to 6D supergravity coming from the supergravity multiplet itself,
which typically contain the dilatini, $\chi$, and/or the
extra-dimensional components of the gravitino, $\psi_m$, $m =
4,5$. If the fermions, $N^I$, in the bulk-brane mixing terms are
linear combinations of these fields, their bulk interactions are
strongly constrained by the fact that they are related by
supersymmetry to the extra-dimensional graviton. In particular,
such fermions cannot have any explicit 6D mass terms at all.
(Supersymmetry similarly constrains the properties of other bulk
fermions, which appear in 6D matter multiplets (hyperini or
gaugini), since such fermions are also typically tied by
supersymmetry to massless bosons in the bulk.

\subsection{The 4D Neutrino Mass Matrix}

We now write down the neutrino bulk-brane mixing which results in
a basis of fields for which we assume the charged-lepton masses to
already be diagonal. Because no bulk 6D mass terms are allowed by
supersymmetry, the complete 4D neutrino mass matrix consists of
eq.~\pref{bulkbranenumixing} plus the Kaluza-Klein masses coming
from the extra-dimensional kinetic terms of the bulk fermions. The
result is
\beq
    S_{\nu \, {\rm mass}} = - \int d^4x \; \sum_{\ell I} \left[
    \frac{\lambda^{(+)}_{aI} v}{r}
    (\nu_a \gamma_\ssl n^I_{\ell+})
    + \frac{\lambda^{(-)}_{aI} v}{r}
    (\nu_a \gamma_\ssl \tilde{n}^{I}_{\ell-})
    +  \frac{c^I_{\ell}}{r} \, (n^{I}_{\ell+} \gamma_\ssl
    \tilde{n}^{I}_{\ell-} ) + \hbox{c.c.} \right] \,,
\eeq
where $v = 246$ GeV is the Standard Model Higgs {\it v.e.v.}, and
$c^I_{\ell} \ne 0$ is an $O(2\pi)$ dimensionless number whose
value depends on the detailed shape of the extra dimensions and
which controls the form of the Kaluza Klein masses through
$m^I_{\ell} = c^I_{\ell}/r$. For instance for compactifications on
a square torus (without brane back-reaction) $\ell$ is a pair of
integers, $\ell = (k_1,k_2)$, and $c_\ell = c_{k_1k_2} = 2\pi(k_1
+ ik_2)$. Here also $\tilde{n}$ denotes the 4D left-handed spinor
which is the conjugate to the right-handed spinor $n$. With (2,0)
--- or higher --- 6D supersymmetry in mind, we do not write an $s = \pm$
dependence for $c_\ell^I$ because we assume both 2D chiralities to
share the same 6D kinetic terms and boundary conditions in the
bulk.

Allowing the indices to run over the ranges $a = 1,2,3$, $s = \pm$
and $I = 1,\dots, n$, the resulting (left-handed) neutrino mass
matrix becomes
\begin{equation} \label{4DMassMatrix}
    {M_\nu}={1 \over r}\left( \begin{array}{cccccccccc}
    0 & 0 & 0 & g_{11}^{(+)} & g_{11}^{(-)} & g_{12}^{(+)} &
    g_{12}^{(-)}
    & g_{13}^{(+)} & g_{13}^{(-)}  & \cdots \\
    0 & 0 & 0 & g_{21}^{(+)} & g_{21}^{(-)}
    & g_{22}^{(+)} & g_{22}^{(-)}
    & g_{23}^{(+)} & g_{23}^{(-)}  & \cdots \\
    0 & 0 & 0 & g_{31}^{(+)} & g_{31}^{(-)}
    & g_{32}^{(+)} & g_{32}^{(-)}
    & g_{33}^{(+)} & g_{33}^{(-)}  & \cdots \\
    g_{11}^{(+)} & g_{21}^{(+)} & g_{31}^{(+)}
    & 0 & c_\ell^{I=1} & 0 &
    0 & 0 & 0 & \cdots \\
    g_{11}^{(-)} & g_{21}^{(-)} & g_{31}^{(-)}
    & c_\ell^{I=1} & 0 & 0 & 0 &
    0 & 0 & \cdots \\
    g_{12}^{(+)} & g_{22}^{(+)} & g_{32}^{(+)}
    & 0 &  0 & 0 & c_\ell^{I=2} & 0 & 0 & \cdots \\
    g_{12}^{(-)} & g_{22}^{(-)} & g_{32}^{(-)}
    & 0 &  0 & c_\ell^{I=2} & 0 & 0 &
    0 & \cdots \\
    \vdots & \vdots & \vdots & \vdots & \vdots & \vdots & \vdots
    & \vdots & \vdots & \ddots
        \end{array} \right)
\end{equation}
where $g_{aI}^{(s)} = \lambda_{aI}^{(s)}v$. For each $\ell$ the
matrix of diagonal $2 \times 2$ blocks with elements $c^{I}_\ell$
is a $2n \times 2n$ matrix corresponding to the $n$ bulk fermions,
assuming these all share the same boundary conditions (and so also
the same KK levels $\ell$). We do not require detailed expresions
for the $c_\ell^I$'s, but because of the brane back-reaction we do
take all of the $c_\ell^I \ne 0$.

In general this matrix is complex and symmetric, and so may always
be diagonalized by finding an appropriate unitary matrix, $U_\nu$,
such that
\begin{eqnarray}
    U^{\sst}_\nu M_\nu \, U_\nu = M^{diag}_\nu = \left(
    {\begin{array}{*{20}c}
   {\mu_1 } & 0 & 0 & \cdots \\
   0 & {\mu_2 } & 0 & \cdots \\
   0 & 0 & \mu_3 & \cdots \\
   \vdots & \vdots & \vdots & \ddots \\
    \end{array}} \right).
\end{eqnarray}
where the superscript `$T$' denotes transposition ({\it not}
hermitian conjugation). The required $U_\nu$ is
\begin{eqnarray}
    U_\nu = V_\nu K \label{Eq:PMNS}
\end{eqnarray}
where $V_\nu$ is the unitary matrix which diagonalizes the
hermitian matrix $M_\nu^\dagger M_\nu$ --- but in the usual
fashion of a similarity transformation: $V_\nu^\dagger \,
(M_\nu^\dagger M_\nu )\, V_\nu = \hbox{diag}(\mu_1^2,\dots)$ ---
and $K$ is a diagonal matrix of phases which can be chosen to
ensure that the $\mu_i$ are non-negative. If the matrix $M_\nu$
should be real --- as we assume for simplicity below --- then
$V_\nu$ can be chosen more simply to be the usual orthogonal
matrix which diagonalizes it. (This diagonalization is performed
explicitly for various choices for the mass matrix below.)

Given such a diagonalization, the three species of 3-brane
neutrino states may be written as a linear combination of the mass
eigenstates according to
\begin{equation}
    \begin{pmatrix} \nu_e \\ \nu_\mu \\ \nu_\tau \end{pmatrix}
    = U \, \begin{pmatrix} \nu_1 \\ \nu_2 \\ \nu_3 \\ \nu_4 \\
    \vdots \end{pmatrix} \,,
\end{equation}
where the non-square matrix $U$ consists of the first 3 rows of
the matrix $U_\nu$. Putting this into the charged-current
interaction, eq.~\pref{CCInteraction}, shows that it is this
matrix, $U$, which plays the role of the PMNS matrix in
extra-dimensional models, and in particular the absence of
evidence for sterile neutrinos requires its left-most $3 \times 3$
block to agree with the $3 \times 3$ PMNS matrix described above,
which is inferred from neutrino-oscillation data. Similarly, its
4th and higher columns must all be sufficiently small not to
conflict with the various bounds on the existence of sterile
neutrinos.

\section{Explicit Models}

In order to proceed further it is necessary to make some choices
for the brane-bulk couplings $\lambda_{aI}^{(s)}$, and for the
shape of the internal dimensions (which determines the
coefficients $c_\ell^I$). In this section we consider the simplest
choices in some detail, guided by considerations of technical
naturalness. For simplicity we assume only a single bulk field
couples to the brane, and so henceforth suppress the bulk index
$I$, and so {\it eg.} $\lambda_{aI}^{(s)} \to \lambda_a^{(s)}$.

\subsection{A Toy Test Case}
Before laying out the couplings which we expect MSLED to provide
and which furnish acceptable phenomenology, it is instructive to
pause here to examine a simple choice for couplings which does
{\it not} work, in order to identify better the issues which must
be addressed by a successful choice. For this purpose consider the
choice where all of the couplings are equal, regardless of flavour
and chirality: $\lambda_{a}^{(s)} = \lambda$ for all $a = 1,2,3$
and $s = \pm$. As is easily verified, with this choice the mass
matrix, eq.~\pref{4DMassMatrix}, leads to the following neutrino
masses and mass eigenstates:

\smallskip\noindent$\bullet$ {\bf Massless States:} Two
massless states are given by
\beqa
    \nu_0 = \frac{1}{\sqrt6} \Bigl( -2 , 1, 1, 0, \cdots
    \Bigr)^T
    \qquad \hbox{and} \qquad
    \nu_0' = \frac{1}{\sqrt2} \Bigl( 0, 1, -1, 0, \cdots
    \Bigr)^T    \,.
\eeqa
Because the nonzero entries only appear in the first three
positions, these massless states do not mix with the infinite
tower of bulk modes.

\smallskip\noindent$\bullet$ {\bf Massive States -- Part I:} For
each $\ell$ there is a neutrino eigenstate having mass $\mu =
|m_\ell| = |c_\ell|/r$, given by
\beq
    \nu_k = \frac{1}{\sqrt2} \Bigl( 0,0,0,\cdots,0,1,-1,0,
    \cdots \Bigr)^T  \,,
\eeq
where $k=1,2,3,...$ and the two nonzero values appear in the
$(2k+2)$'th and $(2k+3)$'th entries of the column vector. These
states clearly have the same KK masses as would have been present
without the brane-bulk mixing term, and since the first three
entries of the eigenstate vanish these states do not mix with the
brane neutrinos.

\smallskip\noindent$\bullet$ {\bf Massive States -- Part II:} All
of the other eigenstates also have masses which are nonzero and
depend nontrivially on the coupling $g = \lambda v$. These states
have masses, $\mu_l = |z_l|/r$, where $z_l$ are defined as the
roots of the following equation
\beq
    z = -6 g^2 \sum_\ell {1 \over c_{\ell} - z} \,.
\eeq
The corresponding eigenstates are given by:
\beq
    \nu_l= N_l  \left( -{1 \over 3 g}, -{1 \over 3 g},
    -{1 \over 3 g}, {1 \over c_{\ell=1} - z_l},
    {1 \over c_{\ell=1}- z_l},{1 \over c_{\ell=2}-z_l},
    {1 \over c_{\ell=2}-z_l}, \cdots \right)^T \,,
\eeq
with normalization constant $N_l$.

It is instructive to ask what happens to these states in the limit
$g \to 0$. As may be seen, for instance by solving this equation
graphically, all but one of the $\mu_l$'s approaches the KK
masses, $|m_\ell| = |c_\ell|/r$ in this limit. The one state whose
mass does not approach $|m_\ell|$ as $g \to 0$ has a mass,
$\mu_0$, which is bounded above by
\beq
    \mu_0 \le \frac{6g^2}{r} \left| {\cal S} \right| \,,
\eeq
where
\beq \label{SSum}
    {\cal S} \equiv \sum_\ell \frac{1}{c_\ell} \,.
\eeq
Evaluating $\mu_0$ perturbatively in $g$ shows that $\mu_0$
actually equals this upper bound up to corrections which are
$O(g^4)$.

Although this mass nominally vanishes as $g \to 0$, what is
interesting is that the sum appearing in $\mu_0$ generically
diverges in the UV, with a divergence linear in the cutoff $L \sim
M_g/m_{\KK} \sim M_g r/(2\pi) \sim 10^{14}$ for two extra
dimensions --- where the numerical estimate takes $M_g \sim 10$
TeV and $m_{\KK} \sim 0.1$ eV. Keeping track of factors of $2\pi$,
and recalling $c_\ell \sim 2 \pi$, we take $\cS \sim
\frac{1}{2\pi} \,( 2\pi L) = L$. This illustrates how
extra-dimensional models can complicate the problem of identifying
masses and mixings in the small-$g$ limit, because the divergences
associated with sums over the very large number of KK modes (like
$\cS$) can overpower small pre-multiplying factors (like $g^2$),
even for the very small values $g^2/(2\pi) \lsim 10^{-8}$ of
interest in what follows.

The divergence appearing in $\mu_0$ may be absorbed as a
renormalization of a counter-term localized on the brane, which
has the form of the usual Standard Model dimension-5 neutrino
mass:
\beq \label{BraneNeutrinoMass}
    S_{\rm brane} = h_{ab} \int d^4x \; (L_a H)(L_b H) \,.
\eeq
The required renormalization is of order $\delta h \sim \lambda^2
M_g \sim 1/M_g$. Since the neutrino masses induced by such an
interaction are of order $hv^2$ and the natural scale for the
$h_{ab}$ is of order $1/M_g$, the existence of such a term
underlines the fact that any successful explanation of neutrino
masses using brane-bulk couplings within MSLED must also explain
the {\it absence} of this pure brane term, which can produce much
too large a mass \cite{Rabi}.

A second type of divergent sum which we shall also encounter is
$g^2 \, \cP$, where
\beq \label{PSum}
    \cP = \sum_\ell \frac{1}{c_\ell^2} \,.
\eeq
Because this diverges logarithmically for two extra dimensions ---
$\cP \sim \frac{1}{(2\pi)^2} (2\pi \ln L) \sim \frac{1}{2\pi} \,
\ln(M_g/m_{\KK}) \sim 32/(2\pi)$ --- the combination $g^2 \cP$ can
be large provided $g^2/(2\pi) \gsim 0.03$. In practice, our
interest in the regime $g \lsim 10^{-4}$ allows us to neglect such
sums, but we nevertheless provide in an appendix expressions for
the eigenvalues and eigenvectors which do not assume the quantity
$g^2 \cP$ to be small.

\subsection{Perturbative Eigenvalues and Eigenvectors}

As the above example shows, agreement with the neutrino data
requires more complicated bulk-brane couplings than those which
were entertained for the Toy Model. Before turning to specific
proposals for more phenomenologically successful brane-bulk
couplings, we pause here to record general expressions for the
eigenvalues and eigenvectors for more complicated neutrino mass
matrices. Keeping in mind the strong bounds on active-sterile
mixing, and in order to keep the analysis simple, we solve the
eigenvalue and eigenvector problem to leading nontrivial order in
the brane-bulk couplings, after first stating some exact
definitions. (We also provide a more detailed calculations for
some phenomenologically interesting choices for the mass matrix in
the appendix.)

Our interest is in diagonalizing the mass matrix of
eq.~\pref{4DMassMatrix}, in the special case of a single bulk
field but with couplings $\lambda^{(\pm)}_a$ which are otherwise
arbitrary. To this end we write the mass eigenstate corresponding
to the lowest three mass eigenvalues, $\mu_i = \mu_{0,\pm} = z_i/r
= O(g^2/r)$ as follows:
\beq \label{Eigenstates}
    \nu_i = N_i \Bigl( n_i e_{1i}, n_i e_{2i}, n_i e_{3i},
    x_{1i}, y_{1i},\dots,x_{\ell i}, y_{\ell i},\dots
    \Bigr)^\sst \,,
\eeq
where $N_i$ is an overall normalization constant and $n_i$ is
chosen to ensure that the $e_{ai}$ satisfy $e_{1i}^2 + e_{2i}^2 +
e_{3i}^2 = 1$. The normalization constant is given by $N_i =
[n_i^2 + \Sigma_i^2]^{-1/2}$ where $\Sigma_i$ denotes the sum
\beq
    \Sigma_i^2 = \sum_\ell (x_{\ell i}^2 + y_{\ell i}^2) \,.
\eeq

In analyzing the physical implications of any such an eigenstate
it is convenient to define the active-sterile mixing angle by
$c_{si} = \cos\theta_{si} = N_i \, n_i$, so that the first three
entries for $\nu_i$ become $c_{si} \,\vec e_i$, with $\vec e_i
\cdot \vec e_i = 1$. In terms of $\Sigma_i$ this is may be written
in the useful equivalent form
\beq
    \tan \theta_{si} = \frac{\Sigma_i}{n_i} \,.
\eeq
Notice that the unitarity of the full matrix $U_{ai}$, together
with the unitarity of the $3 \times 3$ matrix whose elements are
$e_{ai}$, implies that this angle is the one which is relevant to
the total incoherent energy loss into sterile neutrinos from a
charged-current reaction involving lepton flavour $\ell_a$:
\beq
    R_a = \sum_{i>3} |U_{ai}|^2 = 1 - \sum_{i=1}^3 |U_{ai}|^2
    = 1 - \sum_{i=1}^3 c^2_{si} \, |e_{ai}|^2
    = \sum_{i=1}^3 s^2_{si} \, |e_{ai}|^2 \,.
\eeq

Writing, as before, $g_a^{(\pm)} = \lambda_a^{(\pm)} v$, it is
fairly simple to compute the mass eigenvalues and eigenstates to
second order in the small quantities $g_a^{(\pm)}$. Within
perturbation theory, the leading-order masses, $\mu_i$, and the
components, $e_{ai}$, for the lightest three states are determined
by diagonalizing the following perturbative correction to the
mass-matrix within the degenerate 3-dimensional massless subspace:
\beq \label{Perturbativemu}
    \mu_{ab} = - \frac{\cS}{r} \,
    \Bigl( g_a^{(+)} g_b^{(-)} + g_b^{(+)} g_a^{(-)} \Bigr)
     \,.
\eeq
This matrix has rank 2, which ensures that one of the three light
neutrinos remains massless to this order. Once the components
$e_{ai}$, are obtained in this way, the leading expressions for
the components $x_{\ell i}$ and $y_{\ell i}$ are similarly given
by the perturbative expressions
\beq
    x_{\ell i} = - \frac{\vec e_i\cdot \vec g}{c_\ell}^{(-)} \quad
    \hbox{and} \quad
    y_{\ell i} = - \frac{\vec e_i \cdot \vec g}{c_\ell}^{(+)} \,,
\eeq
where $\vec e_i \cdot \vec g^{(\pm)} = \sum_{a=1}^3  e_{ai} \,
g^{(\pm)}_a$. Similarly, the quantity $n_i = 1$ up to corrections
which are 2nd order in the $g^{(\pm)}$. To leading order the
active-sterile mixing angle, $\theta_{si}$, for the three neutrino
species whose mass vanishes as $g \to 0$ then becomes
\beq
    \tan^2 \theta_{si} \approx \Sigma_i^2
    = \sum_\ell \Bigl(x^2_{\ell i} + y^2_{\ell i} \Bigr)
    = \Bigl[ (\vec e_i \cdot \vec g^{(+)})^2 + (\vec e_i \cdot
    \vec g^{(-)})^2 \Bigr] \, \cP \,,
\eeq
where $\cP$ is as in eq.~\pref{PSum}.

\subsection{Approximate Symmetries I} \label{sec33}

With these insights in mind, we next explore a more realistic
model of neutrino masses within the SLED framework. We start in
this section with a model having a minimal flavour content, which
can satisfy all of the required phenomenological constraints but
one. Even though it does not succeed in the end, we describe this
model in detail here for several reasons. First, we do so because
it comes so close and illustrates well the tension between the
various constraints which makes constructing a successful model
difficult. Second, it is useful to describe in detail the symmetry
issues which the model raises, since these also play an important
role in the successful model we describe in the next section.

The simplest and most natural choice to make for the brane-bulk
couplings at the TeV scale is to suppose them to be both
flavour-independent and Lepton-number conserving. Lepton number
conservation is necessary because it precludes the appearance of a
brane neutrino mass like eq.~\pref{BraneNeutrinoMass}.
Flavour-independence is a natural expectation since couplings to
the bulk are gravitational, and so the microscopic physics which
underlies them is unlikely to care about the details of the
low-energy flavour physics on the brane.

These symmetries suggest we take the couplings at the TeV scale to
be
\beq \label{ZeroethOrderCouplings}
    \lambda_{a}^{(+)} = 0 \qquad \hbox{and}
    \qquad \lambda_{a}^{(-)} = \lambda \,,
\eeq
for all brane flavours, $a$, which preserves the Standard-Model
lepton number, $L = L_e + L_\mu + L_\tau$, in addition to the
extra-dimensional chirality, $J$. With this choice there are
precisely 3 massless neutrino eigenstates as well as a tower of KK
modes whose Dirac masses are nonzero. The masslessness of the
three lightest states is protected by the lepton number
invariance. Notice that because these states are degenerate in the
leading approximation, their mutual mixings need {\it not} be
small once this mass matrix is perturbed to obtain more accurately
the neutrino masses and eigenstates. We do not have more than 3
such degenerate states --- unlike previous analyses --- because of
our inclusion of the brane back-reaction, which removes any KK
zero modes.

Now we imagine renormalizing these couplings down to the lower
energies relevant to neutrino physics, and we expect that neither
the flavour nor the lepton-number invariances are likely to
survive this process as an exact symmetry of the low-energy
theory. Flavour symmetry is broken at the very least because the
Standard Model brane couplings themselves distinguish amongst
flavours. Similarly, we have seen that within the bulk the role of
lepton number is carried by local Lorentz rotations, $J$, in the
extra dimensions, and these are naturally broken by the background
extra-dimensional geometry, such as by the background zweibein,
${e_m}^a$.\footnote{Notice that ${e_m}^a$ can remain invariant if
the effects of the local Lorentz transformation acting on $a$ is
cancelled by performing a compensating diffeomorphism on the index
$m$. The diagonal transformation which preserves the zweibein can
then be an unbroken symmetry of the full theory if the
compensating diffeomorphism is also an isometry of the metric
$g_{mn} = {e_m}^a \, {e_n}^b \, \delta_{ab}$. For instance, it can
be unbroken if the internal metric is rotationally-invariant (like
for the 2-sphere) but would be broken if not (like for the
2-torus).} The symmetry-breaking scale for this symmetry is then
naturally of order the KK scale, $m_{\KK} \sim 2\pi/r$. Once such
corrections are included into the brane-bulk couplings we expect
eq.~\pref{ZeroethOrderCouplings} to be replaced by
\beq \label{NextOrderCouplings}
    g_{a}^{(+)} = \epsilon_a^{(+)} \qquad \hbox{and}
    \qquad g_{a}^{(-)} = g + \epsilon_a^{(-)} \,,
\eeq
where as before $g = \lambda v$, and the $\epsilon_a^{(\pm)}$ are
of order $m_{\KK}/M_J \sim K m_{\KK}/M_g \sim 2\pi \, K/(M_g r)
\ll g \sim v/M_g$, where $M_J$ denotes the mass scale which
communicates the symmetry-breaking scale to the brane sector, and
$K = M_g/M_J$ is a dimensionless constant which we choose to
obtain an acceptable neutrino phenomenology.\footnote{As discussed
in ref.~\cite{MSLED}, an explicit microscopic realization of the
SLED picture is likely to involve a number of mass scales smaller
than the 6D Planck constant, $M_g$, such as the string scale,
$M_s$, and the KK scale for any dimensions in addition to the 6
discussed here.} Our task now is to infer the neutrino masses and
mixing which are implied by such a coupling choice.

To second order in $g_a^{(\pm)}$ the effective mass matrix,
eq.~\pref{Perturbativemu}, which determines the mass shift of the
three massless unperturbed states becomes
\beq
    \mu_{ab} \approx - \frac{g \, {\cal S}}{r} \,
    \Bigl( \epsilon_a^{(+)} + \epsilon_b^{(+)}  \Bigr)
     \,,
\eeq
where the approximate equality drops terms which are of order
$\epsilon^2$ in comparison with those which are order $\epsilon
g$.

This matrix has rank 2, which ensures that one of the three light
neutrinos remains massless to this order. The other two neutrinos
acquire masses which are generically of order $\epsilon g/r$ times
the divergent sum $\cS$. Keeping in mind that $\epsilon \sim K
m_{\KK}/M_g$ and the sum diverges linearly in the cutoff, which is
of order $L \sim M_g/m_{\KK}$, we see that the nonzero masses
which are generated are of order $\epsilon g\, L/r  \sim gK/r$.
Because of the protection of the approximate lepton number
symmetry --- {\it i.e.} the small size of $\epsilon_a^{(s)}$ ---
these are the right order of magnitude to describe the observed
neutrino masses.\footnote{To the extent that bounds on sterile
neutrinos require us to take $g \lsim 10^{-4}$ we must imagine $K
\sim 10^4$. An explanation for why $K$ should be so large is a
requirement for any explicit SLED compactification which hopes to
describe neutrino phenomenology. We here content ourselves with a
phenomenological approach, showing that neutrino phenomenology
need not conflict with sterile neutrino bounds in SLED models. We
leave for future work the explicit calculation of
symmetry-breaking effects like $K$ in an explicit model.}

Obtaining the precise form for the masses and mixings of the
lightest three states requires making more specific choices for
the $\epsilon_a^{(+)}$'s, and guided by neutrino-oscillation
phenomenology described in earlier sections we choose these to
preserve the $Z_2$ symmetry which interchanges the 2nd and 3rd
generations. In particular we take
\beq
    \epsilon_a^{(+)} = \Bigl( \epsilon_1 , \epsilon_2, \epsilon_2
    \Bigr)^T \,,
\eeq
in which case the one massless eigenvector is the standard one,
\beq
    \nu_0 = \frac{1}{\sqrt2} \, \Bigl( 0, 1, -1, 0, \cdots
    \Bigr)^T
    \,,
\eeq
for which $x_{\ell 0}=y_{\ell 0}=\sin\theta_{s0} = 0$. The two
nonzero eigenvalues which vanish as $g \to 0$ become
\beq \label{ASIMassesII}
    \mu_\pm = \frac{g \, {\cal S}}{r} \,
    \left[(\epsilon_1 + 2 \epsilon_2) \pm
    \Bigl((\epsilon_1 + 2 \epsilon_2)^2 + 2 (\epsilon_1 -
    \epsilon_2)^2 \Bigr)^{1/2} \right]
    \,,
\eeq
where $\cP$ is as defined in eq.~\pref{PSum}. Notice the $\mu_\pm$
are $O(\epsilon_i g \cS/r)$, as required. The corresponding
neutrino eigenstates have
\beq \label{NuEigenvectorsI}
    \vec e_{\pm} = \frac{1}{\sqrt{2 + \alpha_\pm^2}}
    \Bigl(\alpha_\pm,1,1\Bigr)^\sst \,,
\eeq
where
\beq
    \alpha_\pm = \frac{(\epsilon_1 - 2 \epsilon_2) \pm
    \Bigl((\epsilon_1 + 2 \epsilon_2)^2 + 2 (\epsilon_1 -
    \epsilon_2)^2 \Bigr)^{1/2}}{\epsilon_1 + \epsilon_2} \,.
\eeq
It follows that $\theta_{s\pm}$ is given by
\beq
     \tan^2\theta_{s\pm} \approx \left[ \frac{(\alpha_\pm +2)^2}{
     \alpha_\pm^2 + 2}
     \right] \, g^2 \cP \,,
\eeq
where the approximate equality uses $\epsilon_i \ll g$.

Notice that if $\epsilon_1 = \epsilon_2 \equiv \epsilon$, then
$\alpha_+ = 1$ and $\alpha_- = -2$, and so $\tan^2\theta_{s+} = 3
g^2 \cP$ and $\tan^2\theta_{s-} = 0$. This leads to the following
PMNS matrix for the lightest three neutrino states
\begin{equation}
    U \approx
    \left( \begin{array}{ccccc} -\sqrt{2/3} &&  c_{s+}/\sqrt3 &&  0\\
    1/\sqrt6 &&  c_{s+}/\sqrt3 &&  1/\sqrt2 \\
    1/\sqrt6 &&  c_{s+}/\sqrt3 && 1/\sqrt2 \\
    \end{array} \right) \,,
\end{equation}
whose matrix elements agree (in the limit $\theta_{s+} \to 0$) in
magnitude with the tri-bimaximal mixing matrix described earlier.
For these same choices the neutrino mass values of
eq.~\pref{ASIMassesII} become $\mu_- = 0$ and $\mu_+ = 6 \epsilon
g \cS/r$.

The problem with this model is that it does not appear to have a
parameter range for which the neutrino mass spectrum, mixing
angles {\it and} sterile-active mixing angle are all acceptable
for the same choice of parameters. To see this notice that if we
imagine taking $g$ sufficiently small to minimize the
active-sterile mixing, then we expect the observed neutrino
oscillations to be obtained from the mixings of the three states
whose masses vanish in the limit $g \to 0$. Since our conventions
assume that for three-neutrino mixing it is $\nu_1$ and $\nu_2$
which dominantly participate in solar-neutrino oscillations, we
see that the observed hierarchy $|\Delta m_\odot^2| \ll |\Delta
m_{\rm atm}^2|$ then requires us to take $(\mu_+ - \mu_-)^2 \ll
|\mu_+ \mu_-|$, and so $|\epsilon_1 + 2\epsilon_2| \ll
2|\epsilon_1 - \epsilon_2|$. Notice that this condition is
precisely the opposite to the choice $\epsilon_1 = \epsilon_2$
which led to the successful PMNS mixing matrix, above.

Although we have not found a choice for the $\epsilon_i$ which
produces acceptable masses and mixings using the perturbative
analysis just described, we {\it can} do so if we make $g$ large
enough that the quantity $g^2 \cP$ is not small. In this case the
more exact diagonalization of the appendix shows that
eq.~\pref{ASIMassesII} generalizes to $\mu_\pm = |z_\pm|/r$, with
\beq
    z_\pm =
    g \, {\cal S}
    \, \left[
    \frac{(\epsilon_1 + 2 \epsilon_2) \pm
    \Bigl((\epsilon_1 + 2 \epsilon_2)^2 + 2 (\epsilon_1 -
    \epsilon_2)^2 (1 + 3 g^2 \cP) \Bigr)^{1/2}}{1 + 3g^2 \cP}
    \right]
    \,,
\eeq
and so the condition for obtaining a small $\nu_1 - \nu_2$ mass
splitting becomes
\beq
    (\epsilon_1 + 2\epsilon_2)^2 \ll
    2(\epsilon_1 - \epsilon_2)^2 (1 + 3 g^2 \cP) \,.
\eeq
When $g^2 \cP \gsim 1$ the quantities which control the mixing
angles become
\beq
    \alpha_i = \frac{2[(g^2 + \epsilon_1\epsilon_2) z_i\hP(z_i)
    +g(\epsilon_1 + \epsilon_2) \hS(z_i)]
    }{z_i - (g^2 + \epsilon_1^2) z_i\hP(z_i) - 2 g \epsilon_1 \hS(z_i)}
    \approx \frac{2[g^2 z_i\hP(z_i)
    +g(\epsilon_1 + \epsilon_2) \hS(z_i)]
    }{z_i - g^2 z_i\hP(z_i) - 2 g \epsilon_1 \hS(z_i)}
\eeq
and
\beqa
    x_{\ell i} &=& \frac{g(\alpha_i+2) c_\ell + (\alpha_i\epsilon_1
    + 2 \epsilon_2) z_i}{z_i^2 - c_\ell^2}
    \approx \frac{g(\alpha_i+2) c_\ell}{z_i^2 - c_\ell^2} \nonumber\\
    y_{\ell i} &=& \frac{g(\alpha_i+2) z_i + (\alpha_i\epsilon_1 + 2 \epsilon_2)
    c_\ell}{z_i^2 - c_\ell^2}
    \approx \frac{g(\alpha_i+2) z_i
    }{z_i^2 - c_\ell^2} \,,
\eeqa
where the approximate equality for $y_{\ell i}$ applies for all
$\ell$ if $z_i$ is not too small, but is not valid for $\ell$ too
close to the cutoff if $z_i \sim O(1)$. Here, as before, $|z_i| =
\mu_i r$, and the functions $\hS(z)$ and $\hP(z)$ are defined by
the sums
\beq
    \hS(z) = \sum_\ell \frac{c_\ell}{z^2 - c_\ell^2} \qquad
    \hbox{and} \qquad
    \hP(z) = \sum_\ell \frac{1}{z^2 - c_\ell^2}
\eeq
(from which we see $\hS(0)=-\cS$ and $\hP(0) = -\cP$, and so in
particular $\epsilon_i g \hS(z) \sim O(1)$). Finally, assuming the
sum over $x_{\ell i}^2 + y_{\ell i}^2$ to be dominated by its UV
divergent part, we have
\beq
    \sin^2 \theta_{si} \approx \left[ \frac{ (\alpha_i+2)^2 }{
    \alpha_i^2+2}
    \right] \, g^2 \cP \,.
\eeq

{}From these expressions we see that $\mu_+$ and $\mu_-$ can be
made sufficiently degenerate without giving up an acceptable PMNS
mixing matrix, $e_{ai}$, provided we take $g^2 \cP$ to be
sufficiently large. (Recall that $\cP \sim 32$ and so $g^2\cP$ can
be taken to be large while still keeping $g$ small enough to not
lose perturbative control over loops involving $g$.) However the
penalty paid in this case is unacceptably large active-sterile
mixing angles, $\theta_{si}$. Although we have not been able to
find a choice of parameters within this class which satisfies all
observational constraints, neither have we been able to completely
exclude that this is possible.

\subsection{Approximate Symmetries II} \label{sec34}

We next turn to a pattern of lepton and flavour symmetries which
can produce acceptable neutrino phenomenology. We present this
model in the spirit of an existence proof that such models can be
possible within the SLED framework.

As before we must demand brane-bulk neutrino couplings which
conserve a lepton number in order not to generate too large
neutrino masses as we integrate out scales between the TeV scale
and the sub-eV scale. In this case, however, we assume that our
couplings at the TeV scale separately conserve the lepton numbers
$L_e$ and $L_\mu + L_\tau$, with couplings
\beq \label{ZeroethOrderCouplingsIIa}
    \lambda_a^{(+)} = \Bigl(0 , \lambda, \lambda \Bigr)^\sst
    \qquad \hbox{and}
    \qquad \lambda_a^{(-)} = 0 \,.
\eeq
With this choice all three neutrino masses vanish, and the
masslessness of the three lightest states is protected by the
lepton number invariance.

We next assume that the renormalization down to the sub-eV scale
breaks lepton number, but in such a way that the combination $L_e
- L_\mu - L_\tau$ is less badly broken than the other
combinations, in which case
\beq \label{ZeroethOrderCouplingsII}
    g_{a}^{(+)} = \lambda_a^{(+)}v = \Bigl(0 , g, g \Bigr)^\sst
    \qquad \hbox{and}
    \qquad g_a^{(-)} = \lambda_{a}^{(-)}v = \Bigl( \epsilon,
    0,0 \Bigr)^\sst \,.
\eeq
As before we take $\epsilon \sim K m_{\KK}/M_g \ll g \ll 1$, and
so the lightest nonzero neutrino masses are of order $\epsilon g
\cS/r$. For simplicity we also assume here the permutation
symmetry which interchanges the 2nd and 3rd generations, although
perturbations about this limit are permissible provided they do
not move the atmospheric neutrino oscillations too far away from
maximal mixing.

With these choices we have the usual massless mass eigenstate
\beq
    \nu_0 = \frac{1}{\sqrt2} \, \Bigl( 0, 1, -1, 0, \cdots
    \Bigr)^\sst
    \,,
\eeq
for which $x_{0j}=y_{0j}=\sin\theta_{s0} = 0$. In this case the
unbroken $L_e - L_\mu - L_\tau$ symmetry ensures that the two
nonzero eigenvalues which vanish as $g \to 0$ form a Dirac pair
with lowest-order mass
\beq \label{ASIMassesI}
    \mu_\pm^0 = \frac{\sqrt2 \, \epsilon g \cS}{r}
    \,,
\eeq
and eigenvectors
\beq
    \nu_\pm = \left( \pm \frac{c_{s\pm}}{\sqrt2} , \frac{c_{s\pm}}{2},
    \frac{c_{s\pm}}{2} ,
    x_{1\pm}, y_{1\pm}, \dots
    x_{\ell\pm}, y_{\ell\pm}, \dots \right)^\sst
\eeq
to leading order in $g$. These expressions use the lowest-order
result $\alpha_\pm^0 = \pm \sqrt2$ and
\beq
    x_{\ell\pm} = \mp \frac{\epsilon}{\sqrt2 c_\ell} \quad \hbox{and}
    \quad y_{\ell\pm} = - \frac{g }{c_\ell} \,,
\eeq
and so $\theta_{s+} = \theta_{s-} = \theta_s$, where as before
\beq
    \tan^2\theta_{s} \approx g^2 \cP \,.
\eeq

Phenomenologically acceptable masses and mixings are obtained by
perturbing away from this limit, replacing
eqs.~\pref{ZeroethOrderCouplingsII} with
\beq \label{ZeroethOrderCouplingsIIb}
    g_{a}^{(+)} = \lambda_a^{(+)}v = \Bigl(-2 g' , g, g \Bigr)^\sst
    \qquad \hbox{and}
    \qquad g_a^{(-)} = \lambda_{a}^{(-)}v = \Bigl( \epsilon,
    \epsilon' ,\epsilon' \Bigr)^\sst \,,
\eeq
where $g'/g$ and $\epsilon'/\epsilon$ are both 10\%, but $g'/g -
\epsilon'/\epsilon \lsim 1$\%. This leads to the lowest-order mass
matrix
\beq
    \mu_{ab} = - \frac{{\cal S}}{r} \,
    \Bigl( g_a^{(+)} g_b^{(-)} + g_a^{(-)} g_b^{(+)}  \Bigr)
    \approx - \frac{{\cal S}}{r} \,
    \left( \begin{array}{ccc} -4 \epsilon  g'
    &  \epsilon g -2 \epsilon' g'
    &  \epsilon g -2 \epsilon' g' \\
    \epsilon g -2 \epsilon' g' &  2g \epsilon' &  2g\epsilon' \\
    \epsilon g -2 \epsilon' g' &  2g \epsilon' & 2g\epsilon' \\
    \end{array} \right) \,,
\eeq
whose properties we evaluate perturbatively in $g'/g$ and
$\epsilon'/\epsilon$.

With these choices the massless state is not perturbed: $\delta
\mu_0 = 0$ and $\delta \nu_0 = 0$. On the other hand, including
the leading and next-to-leading shift in the lightest (but
nonzero) two perturbed mass eigenvalues gives
\beq
    \mu_{\pm}=\mu_{\pm}^0 \left[1 \pm
    \sqrt{2}\left(\frac{\epsilon'}{\epsilon}-\frac{g'}{g}\right)
    +\left(\frac{\epsilon'}{\epsilon}\right)^2+
    \left(\frac{g'}{g}\right)^2 + \cdots \right] \,,
\eeq
where $\mu_\pm^0$ is given in eq.~\pref{ASIMassesI}. We work to
next-to-leading order in this expression since our assumptions
imply that the difference $g'/g - \epsilon'/\epsilon$ is the same
order as $(g'/g)^2$ and $(\epsilon'/\epsilon)^2$.

\DOUBLEFIGURE[h]{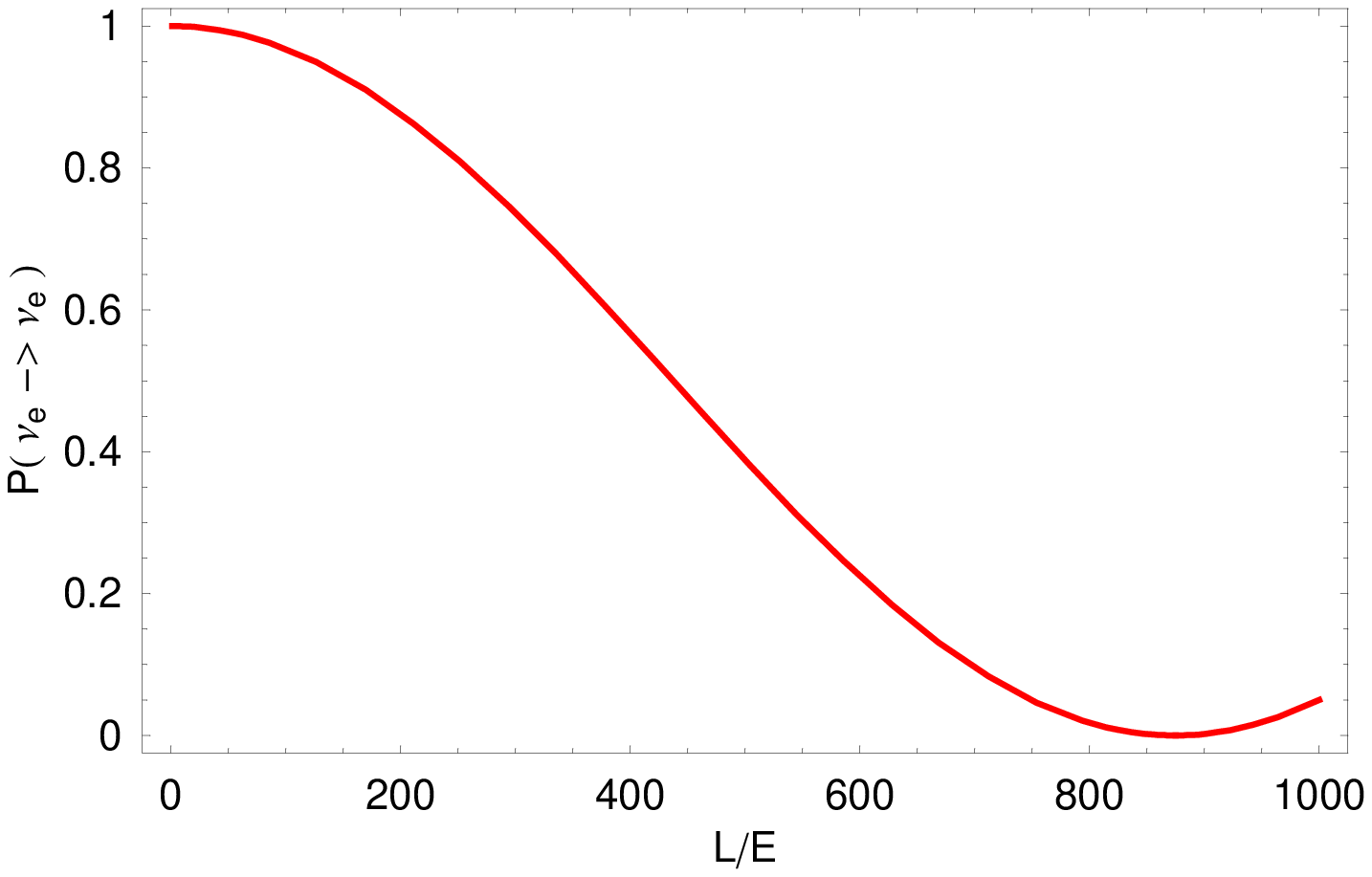, width=.46\textwidth} {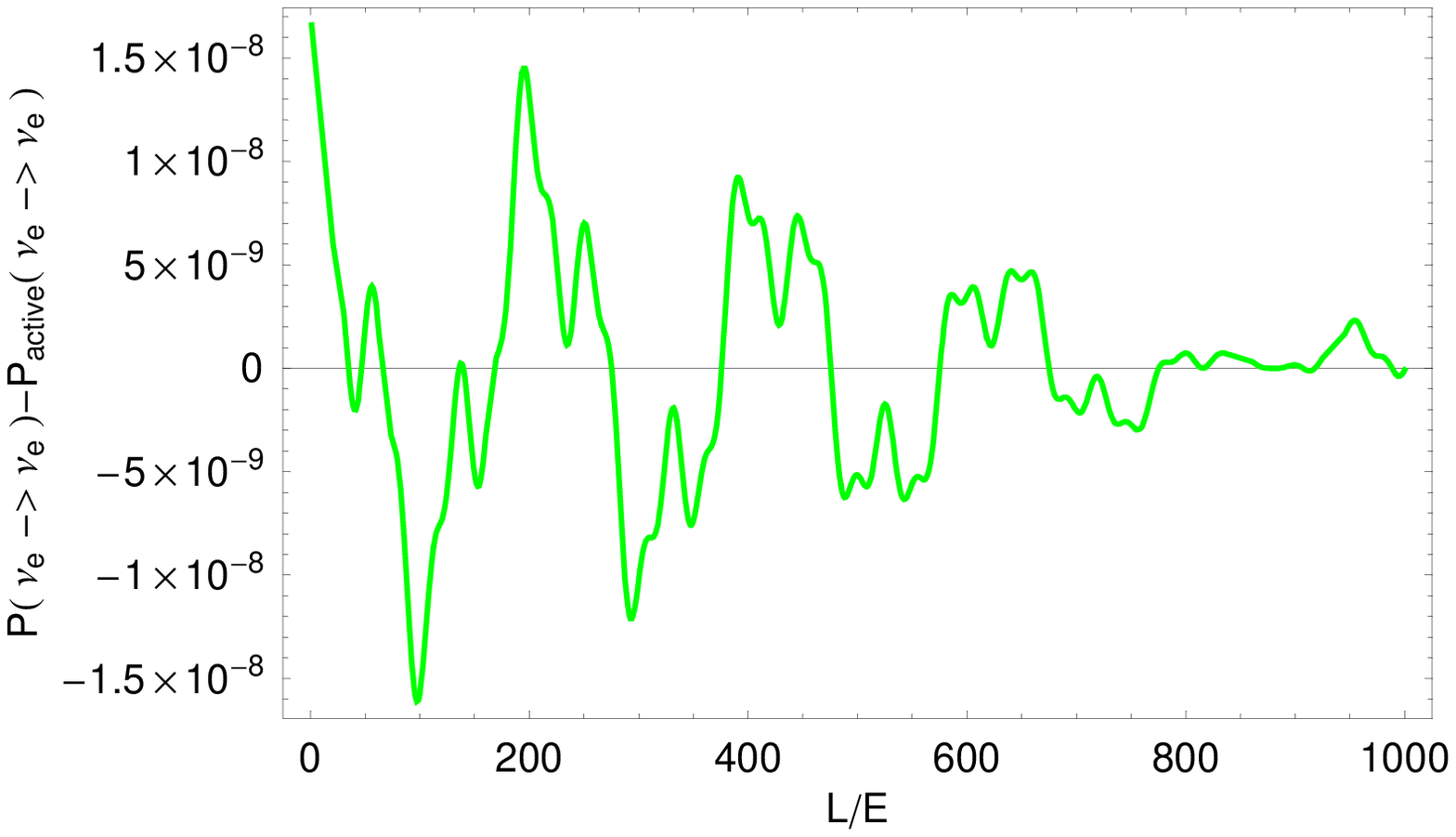,
width=.54\textwidth}{The $\nu_e$ survival probability as a
function of $L/E$ in m/MeV, using $\epsilon'/\epsilon = 0.1$,
$g'/g = 0.11$, $g = 10^{-3}$, ${\cal P}=32$, $1/r = 0.01$ eV and
$\mu_\pm^0 = 0.05$ eV.}{The difference between the full survival
probability and the survival probability computed only using the
`active' states $\nu_0$ and $\nu_\pm$.}

The corrected values for $\alpha_\pm$ are $\alpha_\pm =
\alpha^0_\pm - \delta + \cdots$ where $\alpha^0_\pm = \pm \sqrt2$
and
\beq
    \delta  = 2 \left(
    \frac{\epsilon'}{\epsilon}+\frac{g'}{g} \right) \,,
\eeq
and so the mixing matrix for the lightest three neutrinos becomes
to this order
\begin{equation} \label{UIIMatrix}
    U \approx
    \left( \begin{array}{ccccc} c_{s}(-1/\sqrt2 - \delta/4)
    && c_{s} (1/\sqrt2 - \delta/4) &&  0\\
    c_s (1/2 - \delta/4\sqrt2) &&  c_{s} (1/2 + \delta/4\sqrt2) &&  1/\sqrt2 \\
    c_s (1/2 - \delta/4\sqrt2) &&  c_{s} (1/2 + \delta/4\sqrt2) && -1/\sqrt2 \\
    \end{array} \right) \,.
\end{equation}

Figure 1 shows the survival probability for vacuum oscillations,
computed by numerically performing the mode sum in the expression
\beq
    P(\nu_a \to \nu_a) = \sum_i \Bigl| U_{ai} \Bigr|^4 + 2
    \sum_{i>j} \Bigl| U_{ai} \Bigr|^2 \Bigl| U_{aj} \Bigr|^2 \,
    \cos \left( \frac{\Delta m^2_{ij} L}{2E} \right) \,,
\eeq
using a representative set of parameter values and the KK spectrum
of a square torus (with the zero mode removed). By contrast,
Figure 2 shows the contribution of the sterile neutrinos to the
survival probability. This shows that the oscillations are well
described by oscillations among the usual three active flavours,
with the sterile oscillations only introducing an unobservably
small `fine structure'. Furthermore, the active mixing angles
given by eq.~\pref{UIIMatrix} can describe the data when $g'/g$
and $\epsilon'/\epsilon$ are approximately 10\% in size, because a
10\% shift is sufficient to make the solar mixing angle acceptably
far from maximal mixing. Yet the ratio of mass splittings relevant
to solar and atmospheric neutrinos is also acceptable because
\beq
    \frac{\Delta m^2_\odot}{\Delta m^2_{\rm atm}} = \frac{(\mu_+^2 -
    \mu_-^2)}{\left[ \frac12 (\mu_\pm + \mu_-) \right]^2}
    = \frac{4(\mu_+ - \mu_-)}{\mu_+ + \mu_-} \approx 8 \sqrt2 \left(
    \frac{\epsilon'}{\epsilon} - \frac{g'}{g} \right) \sim 1\% \,.
\eeq
Because of the factor of $8\sqrt2$ this works provided the
difference $\epsilon'/\epsilon - g'/g$ is about 1\% of the value
of each of $\epsilon'/\epsilon$ and $g'/g$ separately, which
unlike the other choices we have made does not seem to follow
purely from symmetry grounds.

Active-sterile mixings are kept acceptable in this model by taking
$g \lsim 10^{-4}$, and this in turn requires at the constant $K$
appearing in $\epsilon = K m_{\KK}/M_g$ is order $10^4$ so that
the mass scales are $\mu_\pm \sim \epsilon g \cS/r \sim K g/r \sim
1/r \sim 0.01$ eV, as required. Since $K = M_g/M_J$, where $M_J$
is the mass scale of the sector which communicates the bulk
lepton-number breaking to the brane, taking $M_g = 10$ TeV implies
the worrisomely-low value $M_J \sim 1$ GeV. The question whether
these choices for $K$, $\epsilon'$ and $g'$ can be obtained from a
simple extra-dimensional geometry is presently under study.

\section{Conclusions}

The coincidence between the values of the recently-discovered
neutrino masses and the Dark Energy density suggests the
possibility of a deep connection between these quantities --- a
possibility which can be more sharply posed within the SLED
proposal for resolving the cosmological constant problem.
Embedding extra-dimensional neutrino models into SLED opens up
many attractive features, largely because the supersymmetry of the
bulk physics necessarily constrains the otherwise arbitrary
properties of the bulk sterile neutrinos. In particular, it
removes the need for the {\it ad hoc} assumption of vanishing bulk
fermion masses, since supersymmetry requires these masses to
vanish. As is also true for non-supersymmetric models, brane-bulk
neutrino mixing provides a natural extension of lepton number to
bulk fermions, by embedding it within local Lorentz rotations of
the large extra dimensions.

In this paper we provide an example of a model which realizes this
connection in a concrete way, thereby motivating a more detailed
examination of neutrino models within the SLED context. The
possibility of so doing comes as a surprise, since past experience
of neutrino models involving large extra dimensions were driven by
sterile-neutrino constraints towards extra dimensions having
properties which are inconsistent with the requirements of the
SLED picture for the cosmological constant problem. In particular,
constraints on catastrophic neutrino emission into the bulk
typically drive model builders to choose only one dimension to
exist near the eV scale, since otherwise the higher-dimensional
phase space can make this emission rate much too large.

By contrast, we here find that astrophysical and cosmological
bulk-neutrino emission can be acceptably small provided that the
brane-bulk mixing is itself small ($g/2\pi \lsim 10^{-4})$. The
somewhat surprising result that this is possible follows largely
from the observation that back-reaction of the stress-energy of
the brane typically removes the massless fermion KK mode. (For
example, many co-dimension 2 branes induce a conical singularity
in the extra-dimensional geometry at the brane position and the
boundary conditions which this conical defect imposes typically
removes the massless bulk fermion modes.) The removal of these
zero modes is crucial for successful neutrino phenomenology, since
even a small brane-bulk couplings typically induces large mixings
amongst any degenerate massless bulk and brane states. If no bulk
zero modes exist then active-sterile neutrino mixings remain small
for small brane-bulk couplings, and the large mixings amongst the
active neutrinos can explain the observed neutrino oscillation
pattern.

We construct a model having acceptable phenomenology by assuming a
relatively mild flavour-dependence for the brane-bulk couplings,
and by taking these couplings to be sufficiently small.
Technically natural small neutrino masses are obtained, with
approximate lepton number protecting the small masses of the
lightest neutrinos. This is possible because lepton number extends
into the extra dimensions as a local Lorentz transformation, and
for generic internal geometries it is naturally broken at the
Kaluza-Klein scale. A potentially unpleasant feature of the small
brane-bulk couplings which we assume is that the amount of lepton
number breaking which is necessary to get acceptable neutrino
oscillations appears to require a relatively large coupling of
bulk lepton-number breaking to the brane neutrinos.

Although it is not yet clear whether such couplings can be
generated from real compactifications, we believe the success of
the models described herein to motivate more detailed studies of
the possibilities offered by neutrino model building within the
SLED framework.

\section*{Acknowledgements}

We thank Marco Cirelli, George Fuller and David London for helpful
discussions. C.B.'s research is supported by a grant from NSERC
(Canada), McMaster University and the Killam foundation and J.M.
receives funds from the Ramon y Cajal Program, FPA2002-00748 and
PNL2005-41. C.B. thanks the hospitality of the Aspen Center for
Physics, where this paper was finalized.

\appendix

\section{Diagonalization for Large Mixings}
\label{Diagonalization}

In this appendix we present the exact results for the eigenvectors
and eigenvalues for the mass matrix in the two cases discussed in
the main text.

\subsection{Approximate Symmetries I}

We take here the symmetry choice of Section \ref{sec33} and impose
$\epsilon_i^{-}=0$ for simplicity, implying
$$g^{(+)}={(\epsilon_1,\epsilon_2,\epsilon_2)}^T \quad g^{(-)}={(g,g,g)}^T$$
The eigenvectors and eigenstates for the mass matrix in this case
come in two distinct kinds:

\medskip\noindent$\bullet$
There is an eigenstate whose eigenvalue is exactly zero:
$$\nu_0={1 \over \sqrt{2}}\left(0,1,-1,0,0,0,0,...\right)$$

\smallskip\noindent$\bullet$
All other eigenstates have nonzero eigenvalues given by $\mu_i =
|z_i|/r$, where the quantity $z_i$ is the solution of the
following transcendental equation
\beqa \label{AppA:EigenvalueEqI}
    && \Bigl(-2 g {\cal
    S}'_i \epsilon_1 + \left[1- {\cal P}'_i \left( g^2+ \epsilon_1^2
    \right)\right] z_i \Bigr) \Bigl(-4 g {\cal S}'_i \epsilon_2 +
    \left[ 1-2{\cal P}'_i \left(g^2+\epsilon_2^2 \right) \right]
    z_i \Bigr) \nonumber \\
    && \qquad \qquad \qquad -2 \Bigl( g {\cal S}'_i (\epsilon_1+\epsilon_2)
    + {\cal P}'_i (g^2+
    \epsilon_1 \epsilon_2) z_i \Bigr)^2=0
    \,,
\eeqa
where
\beq \label{PSPrimeDefs}
    {\cal S}'_i=\sum_{\ell} S_{\ell i} \quad \hbox{with} \quad
    { S}'_{\ell i} \equiv  \frac{c_\ell}{z_i^2-c_{\ell}^2} \quad
    \hbox{and} \quad
    {\cal P}'_i=\sum_{\ell} P'_{\ell i} \quad  \hbox{with}
    \quad {P}'_{\ell i}
    \equiv \frac{1}{z_i^2-c_{\ell}^2}
     \,.
\eeq

The corresponding eigenvectors are
\beq \label{ee1}
    \nu_i={ N}_i \left(
    \alpha_i,1,1,x_{1i},y_{1i},x_{2i},y_{2i},
    x_{3i},y_{3i},... \right) \,,
\eeq
where
\beqa \label{eigen1}
    \alpha_i&=&\frac{2 g {\cal S}'_i  \left(\epsilon_1 +
    \epsilon_2 \right) + 2 {\cal P}'_i z_i \left( g^2 + \epsilon_1
    \epsilon_2 \right)}{z_i \left( 1-(g^2 + \epsilon_1^2) {\cal P}'_i
    \right) - 2 g \epsilon_1 {\cal S}'_i } \,, \nonumber \\
    x_{\ell i}&=& g {S}'_{\ell i}  \left(\alpha_i + 2
    \right)+ z_i {P}'_{\ell i} \left(2 \epsilon_2 + \alpha_i \epsilon_1
    \right) \,, \nonumber \\
    y_{\ell i}&=&{S}'_{\ell i} \left(2 \epsilon_2 + \alpha_i \epsilon_1
    \right)+
   g z_i {P}'_{\ell i} \left( \alpha_i+2 \right)
\eeqa
and the overall normalization factor is given in terms of these by
\beq \label{norm}
    N_i= \frac{1}{\sqrt{2+\alpha_i^2+\sum_j
    \left( x_{ji}^2+y_{ji}^2 \right)}} \,.
\eeq

The solutions to the eigenvalue equation \pref{AppA:EigenvalueEqI}
come in the following two different types:
\begin{itemize}
\item Two light eigenvalues that in the limit of $z_\pm \ll
c_\ell^2$ are:
\beq \label{ee2}
    z_\pm = -g {\cal S}  \frac{
    \left[ (\epsilon_1 + 2 \epsilon_2) \pm
    \Bigl((\epsilon_1+2 \epsilon_2)^2 + 2 (\epsilon_1-\epsilon_2)^2
    u_i \Bigr)^{1/2} \right]}{u_i}
\eeq
where
$$u_i=1+(3
    g^2+ \epsilon_1^2+2 \epsilon_2^2) {\cal P} + 2
    (\epsilon_1-\epsilon_2)^2 g^2 {\cal P}^2 \,.
$$
The corresponding eigenvectors are obtained by using
eq.(\ref{ee2}) in eq.(\ref{ee1}).
\item An infinite tower of eigenvalues, $\mu_{\ell\pm} =
|z_{\ell\pm}|/r$ which come in pairs such that $z_{\ell\pm} \to
\pm c_\ell$ as $g \to 0$. This corresponds to a full tower of
Kaluza Klein states whose masses are slightly perturbed away from
the zeroeth-order result $|c_\ell|/r$ by the brane-bulk couplings.
\end{itemize}

\subsection{Approximate Symmetries II}

We now present expressions for the eigenvalues and eigenvectors of
the mass matrix in the case discussed in Section \ref{sec34}. It
is useful here to distinguish the following two subcases, as
discussed in the main text:
\begin{itemize}
\item[a)] Unperturbed case: $g^+={(0,g,g)}^T$ and
$g^-={(\epsilon,0,0)}^T$
\item[b)] Perturbed case: $g^+={(-2 g',g,g)}^T$ and
$g^-={(\epsilon,\epsilon',\epsilon')}^T$
\end{itemize}

\subsubsection*{Case a: Unbroken Lepton Symmetry}

The eigenvectors in this case resemble those found above:

\medskip\noindent$\bullet$ There is an eigenstate with zero eigenvalue:
$$\nu_0={1 \over \sqrt{2}} \left(0,1,-1,0,0,0,0,...\right) \,.$$

\smallskip\noindent$\bullet$ The rest of the eigenvectors have the form:
$$\nu_i={N_i} \left( \alpha_i,1,1,x_{1i},y_{1i},x_{2i},
y_{2i},x_{3i},y_{3i},... \right)$$
where
\beqa
    \alpha_i&=&\frac{2 {\cal S}'_i g \epsilon }{z_i \left( 1-
    \epsilon^2 {\cal P}'_i
    \right) }, \nonumber \\
    x_{\ell i}&=&{S}'_{\ell i} \alpha_i \epsilon + z_i {P}'_{\ell i} 2
    g , \nonumber \\
    y_{\ell i}&=&{S}'_{\ell i} 2 g +
    z_i {P}'_{\ell i}  \alpha_i \epsilon
\eeqa
and the normalization factor $N_i$ is defined as in
eq.(\ref{norm}). As before, $z_i$ is related to the mass
eigenvalue $\mu_i$ by $\mu_i = |z_i|/r$, with $z_i$ defined as the
solutions to the eigenvalue equation:
\beqa
    z_i^2 \left( 1-  \epsilon^2 {\cal P}'_i \right)
    \left(1-2 g^2 {\cal P}'_i \right)  - 2 \, {\cal
    S}^{'2}_i g^2 \epsilon^2 =0 \,,
\eeqa
with ${\cal P}'$ and ${\cal S}'$ defined as in
eq.~\pref{PSPrimeDefs}. The solutions to this equation come in two
types:
\begin{itemize}
\item Two light eigenvalues given by
$$z_i= \pm 2 \sqrt{\frac{p_i}{u_i}}$$
with
\beqa
    u_i&=&-2-2(\epsilon^2+2 g^2){\cal
    P}-4 \epsilon^2 g^2 {\cal P}^2 \nonumber \\
    p_i&=&-(\epsilon g)^2 {\cal S}^2
\eeqa
where we have simplified the result as appropriate for those
eigenvalues satisfying $z_i^2 \ll c_\ell^2$. The corresponding
neutrino masses are in this case do not depend on the label $\pm$
since they have the common mass $\mu_{\pm}=|z_{\pm}|/r$.

\item Also in this case we have the tower of eigenstates whose
eigenvalues (for small $g$) are perturbations to $z_{\ell \pm}
\sim \pm c_\ell$.
\end{itemize}

\subsubsection*{Case b: Broken Lepton Symmetry}

Perturbing the above results with the addition of $\epsilon'$ and
$g'$ leads to a phenomenologically viable model. In this case we
get the same structure of eigenvectors and eigenvalues, with the
important difference that now the degeneracy between the lightest
massive neutrinos is broken:

\medskip\noindent$\bullet$ One eigenstate remains massless, with
eigenvector
$$\nu_0={1 \over \sqrt{2}}\left(0,1,-1,0,0,0,0,...\right).$$

\smallskip\noindent$\bullet$ The rest of eigenvectors are given by:
\beq \label{eivec2}
    \nu_i={N_i} \left( \alpha_i,1,1,x_{1i},y_{1i},x_{2i},
    y_{2i},x_{3i},y_{3i},... \right) \,,
\eeq
where
\beqa
    \alpha_i&=&\frac{2 {\cal S}'_i \left(g \epsilon - 2 g'
    \epsilon' \right) + 2 {\cal P}'_i z_i \left(- 2 g g' + \epsilon
    \epsilon'\right)}{z_i \left( 1-(4 g'^2 + \epsilon^2) {\cal P}'_i
    \right) + 4 g' \epsilon {\cal S}'_i }, \nonumber \\
    x_{\ell i}&=&{S}'_{\ell i} \left(\alpha_i \epsilon + 2 \epsilon'
    \right)+ z_i {P}'_{\ell i} \left(2 g -2 \alpha_i g'
    \right), \nonumber \\
    y_{\ell i}&=&{S}'_{\ell i} \left(2 g -2\alpha_i g'  \right)+
    z_i {P}'_{\ell i} \left( \alpha_i \epsilon + 2 \epsilon'
    \right)\,
\eeqa
with the normalization factor defined as in eq. (\ref{norm}). The
eigenvalue equation in this case reads:
\beqa
    &&\Bigl( z_i \left[ 1-(4 g'^2 + \epsilon^2) {\cal P}'_i
    \right] + 4 g' \epsilon {\cal S}'_i \Bigr) \Bigl(z_i \left[1-2
    (g^2+\epsilon'^2) {\cal P}'_i \right] - 4 g \epsilon' {\cal S}_i
    \Bigr) \nonumber \\
    && \qquad\qquad\qquad - 2 \Bigl( {\cal S}'_i \left(g \epsilon -
    2 g' \epsilon' \right) +  {\cal P}'_i z_i \left(- 2 g g' +
    \epsilon \epsilon'\right)\Bigr)^2=0 \,.
\eeqa
The solutions to this equation have the same structure as before.
\begin{itemize}
\item There are two light non-degenerated eigenvalues given by
\beq \label{eival2}
    z_i=\frac{v_i \pm \sqrt{v_i^2+4 u_i
    p_i}}{u_i} \,,
\eeq
with
\beqa
    v_i&=-&4(\epsilon g'-\epsilon' g) {\cal S} \nonumber \\
    u_i&=&-2-2(\epsilon^2+2(\epsilon'^2+g^2+2g'^2)){\cal
    P}-4 (\epsilon g + 2 \epsilon' g')^2 {\cal P}^2 \nonumber \\
    p_i&=&-(\epsilon g + 2 \epsilon'g')^2 {\cal S}^2 \,,
\eeqa
where we simplify by assuming $z_i^2 \ll c_\ell^2$. As usual, the
corresponding neutrino mass is $\mu_{\pm} = |z_{\pm}|/r$. The
eigenvectors are obtained by using eq.(\ref{eival2}) in
eq.(\ref{eivec2}).
\item There is a tower of eigenstates organized in pairs, having
eigenvalues which are perturbations of $z_{\ell \pm} \sim \pm
c_\ell$.
\end{itemize}

\end{document}